\documentclass[a4paper,fleqn,usenatbib]{mnras}

\usepackage[T1]{fontenc}
\usepackage{ae,aecompl}

\usepackage{graphicx}	
\usepackage{amsmath}	
\usepackage{amssymb}	
\usepackage{ulem}
\usepackage{enumitem}
\setlist[enumerate]{
  leftmargin=*
}

\usepackage{newtxtext,newtxmath}

\newcommand{\ltday}{{\operatorname{lt-days}}}

\usepackage{eso-pic}

\AddToShipoutPictureBG*{%
  \AtPageUpperLeft{%
    \hspace{0.75\paperwidth}%
    \raisebox{-3.5\baselineskip}{%
      \makebox[0pt][l]{\textnormal{DES-2022-0716}}
}}}%

\AddToShipoutPictureBG*{%
  \AtPageUpperLeft{%
    \hspace{0.75\paperwidth}%
    \raisebox{-4.5\baselineskip}{%
      \makebox[0pt][l]{\textnormal{FERMILAB-PUB-22-568-PPD}}
}}}%

\title[OzDES Mg II R-L relation]{OzDES Reverberation Mapping Program: Mg II Lags and R$-$L relation}

\author[Yu et al.]{
\parbox{\textwidth}{
\Large
Zhefu Yu,$^{1}$
Paul Martini,$^{1,2,3}$
A.~Penton,$^{4}$
T.~M.~Davis,$^{4,5}$
C.~S.~Kochanek,$^{1,2}$
G.~F.~Lewis,$^{6}$
C.~Lidman,$^{5,7}$
U.~Malik,$^{8}$
R.~Sharp,$^{8}$
B.~E.~Tucker,$^{5,8}$
M.~Aguena,$^{9}$
J.~Annis,$^{10}$
E.~Bertin,$^{11,12}$
S.~Bocquet,$^{13}$
D.~Brooks,$^{14}$
A.~Carnero~Rosell,$^{9,15,16}$
D.~Carollo,$^{5,17}$
M.~Carrasco~Kind,$^{18,19}$
J.~Carretero,$^{20}$
M.~Costanzi,$^{17,21,22}$
L.~N.~da Costa,$^{9}$
M.~E.~S.~Pereira,$^{23}$
J.~De~Vicente,$^{24}$
H.~T.~Diehl,$^{10}$
P.~Doel,$^{14}$
S.~Everett,$^{25}$
I.~Ferrero,$^{26}$
J.~Garc\'ia-Bellido,$^{27}$
M.~Gatti,$^{28}$
D.~W.~Gerdes,$^{29,30}$
D.~Gruen,$^{13}$
R.~A.~Gruendl,$^{18,19}$
J.~Gschwend,$^{9,31}$
G.~Gutierrez,$^{10}$
S.~R.~Hinton,$^{4}$
D.~L.~Hollowood,$^{32}$
K.~Honscheid,$^{2,33}$
D.~J.~James,$^{34}$
K.~Kuehn,$^{7,35}$
J. Mena-Fern{\'a}ndez,$^{24}$
F.~Menanteau,$^{18,19}$
R.~Miquel,$^{20,36}$
B.~Nichol,$^{37}$
F.~Paz-Chinch\'{o}n,$^{19,38}$
A.~Pieres,$^{9,31}$
A.~A.~Plazas~Malag\'on,$^{39}$
M.~Raveri,$^{28}$
A.~K.~Romer,$^{40}$
E.~Sanchez,$^{24}$
V.~Scarpine,$^{10}$
I.~Sevilla-Noarbe,$^{24}$
M.~Smith,$^{41}$
E.~Suchyta,$^{42}$
M.~E.~C.~Swanson,$^{14,34}$
G.~Tarle,$^{30}$
M.~Vincenzi,$^{41,43}$
A.~R.~Walker,$^{44}$
N.~Weaverdyck,$^{30,45}$
\\
{\it \footnotesize (Affiliations listed at the end of the paper)}
}
}

\date{Accepted XXX. Received YYY; in original form ZZZ}

\pubyear{2022}

\begin{document}
\label{firstpage}
\pagerange{\pageref{firstpage}--\pageref{lastpage}}
\maketitle

\begin{abstract}
The correlation between the broad line region radius and continuum luminosity ($R-L$ relation) of active galactic nuclei (AGN) is critical for single-epoch mass estimates of supermassive black holes (SMBHs). At $z \sim 1-2$, where AGN activity peaks, the $R-L$ relation is constrained by the reverberation mapping (RM) lags of the Mg II line. We present 25 Mg II lags from the Australian Dark Energy Survey (OzDES) RM project based on six years of monitoring. We define quantitative criteria to select good lag measurements and verify their reliability with simulations based on both the damped random walk stochastic model and the re-scaled, re-sampled versions of the observed lightcurves of local, well-measured AGN. Our sample significantly increases the number of Mg II lags and extends the $R-L$ relation to higher redshifts and luminosities. The relative iron line strength $\mathcal{R}_{\rm Fe}$ has little impact on the $R-L$ relation. The best-fit Mg II $R-L$ relation has a slope $\alpha = 0.39 \pm 0.08$ with an intrinsic scatter $\sigma_{\rm rl} = 0.15^{+0.03}_{-0.02}$. The slope is consistent with previous measurements and shallower than the H$\beta$ $R-L$ relation. The intrinsic scatter of the new $R-L$ relation is substantially smaller than previous studies and comparable to the intrinsic scatter of the H$\beta$ $R-L$ relation. Our new $R-L$ relation will enable more precise single-epoch mass estimates and SMBH demographic studies at cosmic noon. 
\end{abstract}




\section{Introduction} \label{sec:intro}
Accurate mass measurements of supermassive black holes (SMBHs) are critical for understanding their growth over cosmic time. In the local universe, studies have used spatially resolved kinematics of stars or gas at the centres of galaxies to determine the SMBH mass \citep[e.g.,][]{Kormendy1995,Gebhardt2009,Barth2016}. However, it is difficult to extend this method to higher redshifts due to the angular resolution limits of current facilities. 

The reverberation mapping (RM) technique is a robust method to measure the SMBH mass in active galactic nuclei (AGN) outside the local universe \citep{Blandford1982,Peterson1993}. The AGN broad emission lines vary in response to the stochastic variation of the continuum emission after a time delay due to the light travel time from the central accretion disk to the broad line region (BLR). The time lag $\tau$ is correlated with the SMBH mass through the virial theorem
\begin{equation}
M_{\rm BH} = \frac{f c\tau \Delta v^2}{G}
\label{eq:bhmass}
\end{equation}
where $f$ is a ``virial factor'' determined by the dynamics and structure of the BLR, $R = c\tau$ gives the characteristic BLR size and $\Delta v$ is the velocity width of the broad lines. To measure the time lag, RM campaigns generally monitor the target for months to years to obtain the photometric and spectroscopic lightcurves, which requires substantial observational resources.

An important result of RM studies is a correlation between the BLR radius $R$ and the AGN continuum luminosity $L$. The existence of an $R-L$ correlation enables a measurement of the SMBH mass from just a single-epoch spectrum. Such single-epoch estimates can be applied to large sample of AGN, making them critical for SMBH demographic studies \citep[e.g.,][]{Vestergaard2008,Kelly2013}. The $R-L$ relation has been well-constrained in nearby AGN using time lag measurements of the H$\beta$ line \citep[e.g.,][]{Kaspi2000,Bentz2009,Bentz2013,Grier2017}. However, the H$\beta$ line is not present in optical spectra of higher redshift AGN. At higher redshifts we need $R-L$ relations for other lines, such as Mg II and C IV. Despite previous RM studies of Mg II \citep[e.g.,][]{Metzroth2006,Shen2016,Lira2018,Czerny2019,Homayouni2020,Zajacek2020,Yu2021} and C IV \citep[e.g.,][]{Kaspi2007,Lira2018,Hoormann2019,Grier2019}, the $R-L$ relations of these lines remain poorly constrained. 

Star formation and AGN activity peak at redshifts of $z \sim 1 - 2$ \citep[e.g.,][]{Wolf2003,Ueda2014}. The single-epoch mass estimates at this epoch of cosmic noon are mainly based on the $R-L$ relation of the Mg II line because it is the major broad line observable in optical spectra. Unfortunately, there were less than ten Mg II lags available until recently \citep{Metzroth2006,Shen2016,Lira2018,Czerny2019}. Therefore, early studies of SMBH demographics generally calibrated the Mg II $R-L$ relation to match the H$\beta$ lags instead of using the direct Mg II lags \citep[e.g.,][]{Vestergaard2009}. This could lead to a bias because the collisionally-excited Mg II line may respond differently to the continuum variability than the photoionised Balmer lines \citep[e.g.,][]{Guo2020}. It is therefore critical to better constrain the Mg II $R-L$ relation directly with a larger sample of lag measurements. 

A promising way to quickly increase the number of lag measurements is through RM campaigns with wide-field multi-fibre spectrographs and imaging facilities, such as the Sloan Digital Sky Survey (SDSS) RM project \citep[e.g.,][]{Shen2015} and the Dark Energy Survey (DES) - Australian DES (OzDES) RM project \citep{King2015,Hoormann2019,Yu2021}. Both surveys have monitored $\sim 800$ AGN at redshifts up to $z \sim 4.5$ for $\sim 5-6$ years. The large sample size and long time duration potentially allow these projects to produce a large number of lag measurements. However, these RM campaigns also face significant challenges, most notably due to the complexity of flux calibrating fibre spectra, the relatively low cadence and signal-to-noise ratio (SNR) of the spectroscopy, and the gaps between observing seasons. As a result, the probability distribution of the lag can be complicated and it is non-trivial to properly select reliable lag measurements and define lag uncertainties. 

\citet{Homayouni2020} presented Mg II lags for 57 quasars from the SDSS RM project, from which they identified 24 quasars as their ``gold sample'' with the most reliable lags. They derived a Mg II $R-L$ relation with a slope of 0.31 and an intrinsic scatter of 0.36 dex from this gold sample. Their $R-L$ relation has a shallower slope and larger scatter than the H$\beta$ relations that typically have a slope of $\sim 0.5$ and an intrinsic scatter of $\sim 0.13 - 0.2$ dex \citep[e.g.,][]{Bentz2013,Du2016}. \citet[][hereafter Y21]{Yu2021} presented nine Mg II lags from the first five years of data on about half of the OzDES RM sample. Their results show much less scatter than \citet{Homayouni2020}, but the sample had a small dynamic range in luminosity and therefore could not independently constrain the slope of the $R-L$ relation. 

\citet{Guo2020} used photoionization models to show that the response of the Mg II line to continuum variability was weaker than the Balmer lines, which could potentially explain the scatter of the Mg II $R-L$ relation. However, their model does not quantitatively predict the scatter of the Mg II $R-L$ relation. RM studies of the H$\beta$ line found that AGN with larger Eddington ratios generally had smaller lags at fixed luminosities \citep[e.g.,][]{Du2016,Du2018,Dalla2020}, which may also explain the shallower slope and larger intrinsic scatter of the $R-L$ relation when this effect is not included. \citet{Martinez2020} reduced the scatter of the Mg II $R-L$ relation to $\sim 0.1$ dex by including the Eddington ratio as an additional parameter, but their Eddington ratio estimates depended on the lag measurements. \citet{Khadka2022} used the ratio $\mathcal{R}_{\rm Fe}$ of the iron line flux to the Mg II line flux as an independent indicator of the Eddington ratio and found that it had no significant impact on the scatter of the Mg II $R-L$ relation. A larger Mg II lag sample with a wider redshift and luminosity range is critical for better constraining the Mg II $R-L$ relation and understanding its intrinsic scatter. 

In this paper we present 25 Mg II lags from the full OzDES RM sample with six years of data and derive a new Mg II $R-L$ relation. Our sample is homogeneously defined through lag quality criteria that are verified by multiple simulations and statistical tests. We describe our observations and spectroscopic analysis in Section \ref{sec:obs}. Our time series analysis and lag measurements are discussed in Section \ref{sec:lcanl}. Section \ref{sec:reliability} discusses the reliability assessments of the lag measurements. We present the black hole mass and $R-L$ relation in Section \ref{sec:RL_BHMass}. Section \ref{sec:summary} summarizes the paper. This paper adopts a $\Lambda$CDM cosmology with $H_0 = 70 \, {\rm km/s/Mpc}$, $\Omega_m = 0.3$, $\Omega_{\Lambda} = 0.7$.

\section{Observations and Spectroscopic Analysis} \label{sec:obs}
DES is a 6-yr wide-area photometric survey that began in 2013 \citep{DESDR1}. The survey took images in the $grizY$ bands using the Dark Energy Camera \citep[DECam, ][]{DECam} with a $2.2^{\circ}$ diameter field of view on the 4-m Victor M. Blanco telescope at the Cerro Tololo Inter-American Observatory. In addition to the $5000 \, {\rm deg}^2$ wide-area survey, DES observed 10 supernova (SN) fields approximately every week for the first five years and every three weeks for Y6.  

\begin{figure}
\includegraphics[width=\linewidth]{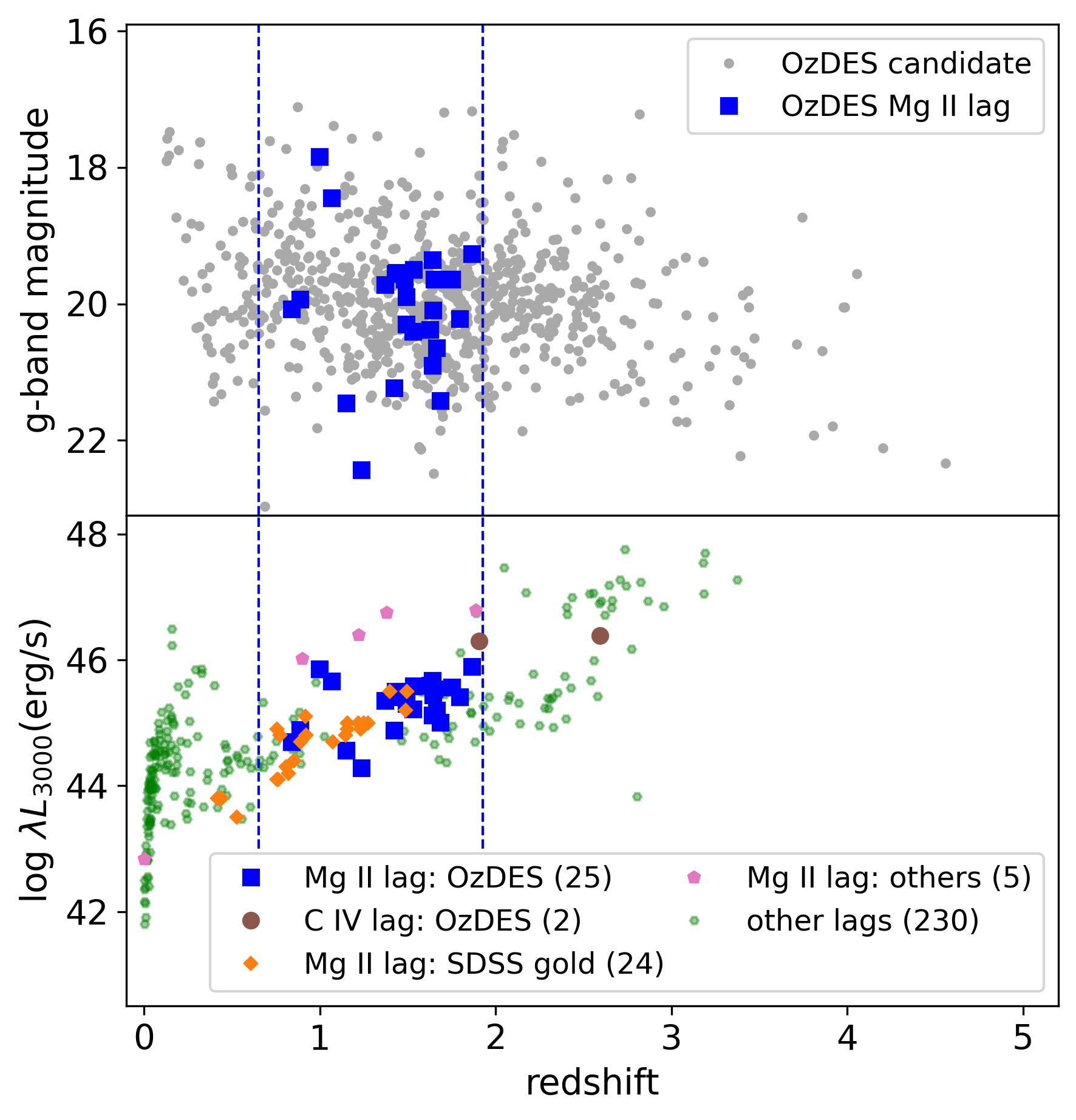}
\caption{({\it Upper panel}) $g$-band apparent magnitude versus redshift of the OzDES RM quasars (grey points) and the sub-sample that has successful Mg II lag measurements (blue squares). The blue dashed lines show the redshift range of quasars that we analyzed in this paper. ({\it Lower panel}) Monochromatic luminosity at 3000 \AA\, versus redshift. The blue squares, brown points, orange diamonds, pink pentagons and green hexagons represent the OzDES Mg II lag sample (this work), the OzDES C IV lag sample \citep{Hoormann2019}, the gold sample of the SDSS Mg II lags \citep{Homayouni2020}, other sources with Mg II lags from literature \citep{Metzroth2006,Lira2018,Czerny2019,Zajacek2020,Zajacek2021} and other RM sources from literature \citep{Peterson2004,Peterson2005,Kaspi2007,Bentz2009,Denney2010,Barth2011_Mrk50,Barth2011_Zw229,Grier2012,Bentz2013,Barth2013,Bentz2014,Du2014,Pei2014,Trevese2014,Wang2014,Du2015,Du2016,Bentz2016,Lu2016,Fausnaugh2017,Grier2017,Du2018,Lira2018,Grier2019,Zhang2019,Hu2021,Li2021,Williams2021,U2022}. The number of lags in each sample is given in the brackets. We converted the monochromatic luminosities at 5100 \AA\, and 1350 \AA\, to 3000 \AA\, using the bolometric corrections from \citet{Richards2006}.}
\label{fig:sample}
\end{figure}

OzDES is a spectroscopic follow-up survey in the DES SN fields that covers roughly the same time baseline as DES \citep[e.g.,][]{Yuan2015,Childress2017,Lidman2020}. The spectra cover the $\sim 3700 - 8900$ \AA\, wavelength range and were taken using the AAOmega spectrograph \citep{AAOmega} with the Two Degree Field (2dF) multi-fibre positioner \citep{2dF} on the 4-m Anglo-Australian Telescope (AAT). The OzDES RM project is one of the key OzDES science projects. Over the six DES observing seasons, it monitored 735 quasars in the DES SN fields with about monthly cadence. An observing season spans about six months from July to January. Figure \ref{fig:sample} shows the apparent magnitude and redshift distribution of the OzDES RM quasars. The sample spans an AB magnitude range of $g \sim 17 - 23$ mag and a redshift range of $z \sim 0.1 - 4.5$. 

We use the pipeline from \citet{Hoormann2019} to calibrate the spectra. The pipeline first calculates the scaling factors from the instrumental flux derived by integrating the extracted and wavelength calibrated spectra within the DES filters to the DES photometric flux in the $gri$ bands. The DES flux is derived from the linear interpolation of the two DES photometric epochs bracketing the spectroscopic epoch. It then fits the scaling factors with a second-order polynomial and generates flux calibrated spectra by multiplying the polynomial to the extracted and wavelength calibrated spectra. 

The OzDES RM project has developed simulation frameworks for survey design and lag quality assessment \citep{King2015,Penton2022,Malik2022} and published early measurements of continuum lags \citep{Mudd17,Yu2020_PhotRM}, Mg II lags \citep{Yu2021} and C IV lags \citep{Hoormann2019} based on the data from the first four to five years. In this paper we analyze all six years of data for 453 quasars at $0.65 < z < 1.92$. We model and subtract the continuum + iron emission in the rest-frame wavelength region spanning $2260 - 3050$ \AA\, near the Mg II line (Section \ref{subsec:ironfit}). The redshift range of our candidate sample ensures that the iron fitting range is fully covered by the OzDES spectra. The spectroscopic calibration could introduce correlated errors between the spectral pixels over a wide wavelength range. We estimate the calibration uncertainties based on the F-stars monitored by OzDES (Section \ref{subsec:calerr}).

\subsection{Continuum \& Iron Subtraction} \label{subsec:ironfit}
A challenge in analyzing the Mg II line is the strong contribution from nearby iron emission lines \citep[e.g.,][]{Wills1980,Wills1985,Verner1999}. The iron lines could also have reverberation signals that contaminate the Mg II lag signal \citep[e.g.,][]{Barth2013}. We fit and subtract the continuum + iron emission and derive the Mg II line flux using the pipeline from Y21. Our model 
\begin{equation}
f_{\rm model}(\lambda) = f_{\rm c}(\lambda) + f_{\rm Fe}(\lambda)
\label{eq:ironmodel_1}
\end{equation}
consists of a power-law continuum
\begin{equation}
f_{\rm c}(\lambda) = A_{\rm c} (\lambda/\lambda_0)^{\alpha}
\label{eq:ironmodel_2}
\end{equation}
with an index $\alpha$ and a normalisation $A_{\rm c}$ at $\lambda_0 = 2599$ \AA\,\footnote{The constant $\lambda_0$ is chosen to reduce the error correlation between the power-law index $\alpha$ and normalisation $A_{\rm c}$ by reducing the non-diagonal terms of the Fisher matrix.}, and an iron emission line component
\begin{equation}
f_{\rm Fe}(\lambda) = A_{\rm t} f_{\rm t}(\lambda) \ast G(w) 
\label{eq:ironmodel_3}
\end{equation}
modeled by an iron template $f_{\rm t}(\lambda)$ of a normalisation $A_{\rm t}$ convolved with a Gaussian kernel $G(w)$ of a width $w$ to account for the velocity broadening of the BLR. The four free parameters of the model are $A_{\rm c}$, $\alpha$, $A_{\rm t}$ and $w$. 

We adopt the empirical iron template from \citet{Vestergaard2001} based on the Seyfert galaxy I Zwicky 1. While the \citet{Vestergaard2001} template did not model the iron emission under the Mg II line, Y21 tested other iron templates by \citet{Tsuzuki2006} and \citet{Salviander2007} which modeled the iron emission under Mg II and found that the choice of iron template had little impact on the lag measurements. We fit the spectra over the rest-frame wavelength ranges 2260 - 2690 \AA\, and 2910 - 3050 \AA\, using a Markov chain Monte-Carlo (MCMC) sampler. We do not include the wavelength range near the Mg II line, since it is difficult to distinguish the iron emission from the strong Mg II line within this range. Using a wider fitting range has little impact on our lag measurements. When fitting the single-epoch spectra, we cannot constrain the broadening width $w$ very well due to the low SNR, so we fix it to the best-fit value found for the co-added spectra. While the width of the iron lines could vary in response to the continuum variability \citep[e.g.,][]{Korista2004,Guo2020,Wang2020}, Y21 found such variability had little impact on the derived Mg II line flux. We do not include Balmer continuum in the fitting. Based on the simulations in \citet{Lawther2018}, we estimate that the variability of the Balmer continuum under the Mg II line is much smaller than the uncertainty of the Mg II flux. Therefore, the Balmer continuum will not have significant impact on our lag measurements.

\begin{figure}
\includegraphics[width=\linewidth]{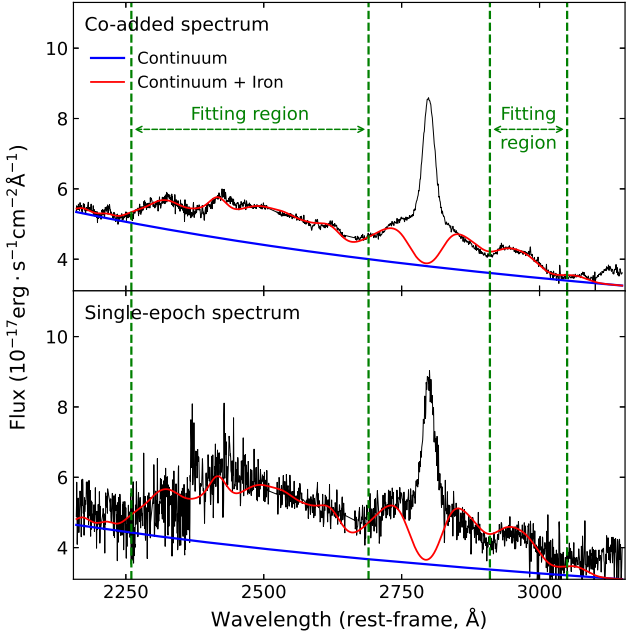}
\caption{Examples of the continuum + iron modeling of the co-added spectrum (upper panel) and a single-epoch spectrum (lower panel) in the rest-frame of DES J003052.76$-$430301.08. The green dashed lines show the regions where we fit the spectra. The blue and red solid lines are the best-fit continuum and continuum + iron models, respectively.}
\label{fig:ironfit}
\end{figure}

We derive the Mg II flux as $F_{\rm line} = F_{\rm total} - F_{\rm model}$, where $F_{\rm total}$ is the integrated spectral flux over the rest-frame wavelength range 2700 - 2900 \AA\, before continuum and iron subtraction and $F_{\rm model}$ is the integrated continuum + iron model flux. The line flux uncertainty is estimated as 
\begin{equation}
\sigma_{\rm line}^2 = \sigma_{\rm total}^2 + \sigma_{\rm model}^2
\end{equation}
\label{eq:ftoterr_1}
where
\begin{equation}
\sigma_{\rm total} = \sqrt{\sum_i \sigma_{{\rm total},i}^2} \Delta \lambda 
\end{equation}
\label{eq:ftoterr_2}
is the uncertainty in the total flux $F_{\rm total}$, $\sigma_{{\rm total},i}$ is the uncertainty of the $i^{\rm th}$ pixel, $\Delta \lambda$ is the wavelength pixel size and $\sigma_{\rm model}$ is the uncertainty in the model flux $F_{\rm model}$. We estimate $\sigma_{\rm model}$ as the scatter of the model fluxes in the MCMC chain. Figure \ref{fig:ironfit} shows an example of the continuum + iron modeling. The model generally matches the observed spectra well. We visually inspect all spectra and exclude epochs from further analysis where the best-fit model fails to match the spectra or where the spectra are contaminated by instrumental artifacts, such as the bump at $\sim 7100$ \AA\, created by an LED in the 2dF gripper gantry. Figures for all spectra of our final sample of quasars are available in the supplementary material. 

\subsection{Calibration Uncertainty} \label{subsec:calerr}
We estimate the calibration uncertainty and propagate it to the error budget of the Mg II line flux using the pipeline from Y21 based on the F-stars observed by OzDES in each epoch. For each F-star, we calculate the mean $\langle f_{*,i} \rangle$ and rms $S_{*,i}$ of the calibrated spectra over all epochs, where $i$ corresponds to the $i^{\rm th}$ pixel. Since F-stars do not have intrinsic variability, the residual variation $S_{*,i}$ is due to a combination of the mostly uncorrelated photon noise and the correlated calibration uncertainty. We cannot directly derive the calibration uncertainty from $S_{*,i}$ since it is dominated by the photon noise at the pixel level. We therefore bin the spectra by 300 \AA\, to suppress the photon noise, while the correlated calibration uncertainty remains. The residual variation of the binned spectra is an estimate of the calibration uncertainty.  

We propagate the calibration uncertainty to the line flux uncertainty using a Monte-Carlo method. We define a ``warping function''\begin{equation}
W_{bj} = f_{*,bj} / \langle f_{*,b} \rangle
\label{eq:warping_func}
\end{equation}
where $f_{*,bj}$ is the flux of the binned F-star spectra in the $b^{\rm th}$ bin of the $j^{\rm th}$ epoch and $\langle f_{*,b} \rangle$ is the mean of the binned spectra over all epochs. We then derive a continuous warping function $W_j(\lambda)$ by interpolating the discrete function $W_{bj}$ with a $3^{\rm rd}$ order spline function. Each epoch of the F-star spectra provides a warping function $W_j(\lambda)$, and each warping function is a realization of the spectral variability due to the calibration uncertainty. We create 2343 warping functions based on 161 F-stars. This provides a ``library'' of warping functions for each spectroscopic epoch. Multiplying the quasar spectrum with the warping functions of the corresponding epoch gives realizations of the quasar spectra warped by the calibration uncertainty. In each warped spectra, we calculate the Mg II flux using the method described in Section \ref{subsec:ironfit}. We then multiply the fractional variation of the warped Mg II fluxes by the Mg II flux of the observed spectra to derive the line flux uncertainty contributed by the calibration procedure. This uncertainty is added in quadrature to the line flux uncertainty estimated in Section \ref{subsec:ironfit}. The median calibration uncertainty is $\sim 4\%$.

The calibration pipeline uses the linear interpolation of the two neighboring DES epochs to estimate the continuum flux at a spectroscopic epoch, which is a reasonable estimate since the cadence of the DES photometric lightcurve is much higher than the OzDES spectroscopy. While the photometric cadence in Y6 is lower than the first five years, the Y6 spectra are generally taken within a few days of the neighboring photometric epochs. The only cases where the linear interpolation can significantly increase the calibration uncertainty are two spectroscopic epochs taken at 2018-06-15 (MJD 58284) and 2018-06-24 (MJD 58293) for some quasars. They are in the seasonal gap between DES Y5 and Y6 and therefore have no nearby photometric epoch. This introduces additional calibration error due to the large uncertainty of the continuum flux. In these two cases we interpolate the photometric lightcurve with a damped random walk (DRW) stochastic process and calculate the fractional uncertainty of the DRW model at the time of the spectroscopic epoch. This characterizes the uncertainty in the overall normalisation of the calibration and we add it as an additional uncertainty to the error budget of the Mg II line flux. The additional uncertainty is $\sim 10\%$ for these two cases.

\section{Time Series Analysis} \label{sec:lcanl}
We use the $g$-band photometry from the DES database for the continuum lightcurves and the pipeline described above to create Mg II line lightcurves. Machine-readable lightcurves for our final sample are available in the supplementary material. We use JAVELIN \citep[e.g.,][]{Zu2011,Zu2013} and the interpolated cross-correlation function \citep[ICCF, e.g.,][]{Gaskell1987,Peterson1998,Peterson2004} method to measure the time lags. JAVELIN uses a DRW stochastic process to interpolate the lightcurve and assumes that the line lightcurve is the continuum lightcurve convolved with a top-hat transfer function. JAVELIN fits the continuum and line lightcurves simultaneously with a MCMC sampler to derive the posterior probability distributions of the amplitude $\sigma_{\rm drw}$ and characteristic time scale $\tau_{\rm drw}$ of the DRW model, the scale $s_l$ and width $w_l$ of the transfer function, and the mean time lag $\tau$. The algorithm sets the prior of the DRW parameters $\sigma_{\rm drw}$ and $\tau_{\rm drw}$ as their posterior distributions from fitting only the continuum lightcurve. We use a flat prior for the time lag $\tau$ within [$\tau_{\rm min}$,$\tau_{\rm max}$] days, while we allow the other parameters to vary freely. 

We use \textsf{PyCCF} \citep{pyccf}, a python implementation of the ICCF method. This method linearly interpolates the lightcurves and calculates the cross-correlation function (CCF) within a lag range [$\tau_{\rm min}$,$\tau_{\rm max}$] days. It then uses the range of the CCF where the cross-correlation coefficient is at least 80\% of the maximum and estimates the lag as its centroid or peak. For the lag uncertainty, it first creates lightcurve realizations by randomizing the single-epoch flux by its uncertainty and randomly sub-sampling the epochs with replacement. It then calculates the centroid and peak of the CCF for each realization to create the cross-correlation centre distribution (CCCD) and the cross-correlation peak distribution (CCPD). The scatter of these distributions gives the lag uncertainty estimate. We generate 8000 realizations and adopt the realizations with $r_{\rm peak}>0.5$ to create the CCCD and CCPD, where the $r_{\rm peak}$ is the peak value of the CCF. We use CCCD as the fiducial lag probability distribution of the ICCF method, since previous work found it provided better lag estimates than CCPD \citep[e.g.,][]{Peterson2004}. 

The lag distributions from JAVELIN and ICCF usually have multiple peaks that are mostly due to the aliasing effects caused by the seasonal gaps, so it is non-trivial to identify successful lag measurements and estimate their uncertainties. We use two different methods to analyze the lag distributions, leading to two sets of criteria to select successful lag measurements.

\begin{table*}
\small
\renewcommand{\arraystretch}{1.25}
\begin{tabular}{ccllcllcccc}
\hline
{ Source Name} & { $z$} & { $\tau_{\rm JAV}$} & { $\tau_{\rm ICCF}$} & { $f_{\rm peak}$} & { FPR} & { $f_{3000}$} & { log($\lambda L_{3000}$ [erg/s])}  & { $\mathcal{R}_{\rm Fe}$}  & { $\sigma_{\rm line}$} & { log($M_{\rm BH}/M_{\odot}$)}\\
 &  & { (days)} & { (days)} &  & { (\%)} & (see caption)  &  &  & { (km/s)} & \\
\hline
DES J024340.09$+$001749.40 & 1.44 & $818^{+39}_{-6}$ & $797^{+54}_{-20}$ & 0.76 & 7.9 & $2.75\pm 0.15$ & $45.41\pm 0.02$ & $1.44\pm 0.17$ & 3181.0 & 9.46 \\
DES J025254.18$+$001119.70 & 1.64 & $415^{+27}_{-166}$ & $420^{+27}_{-44}$ & 0.75 & 3.7 & $0.93\pm 0.02$ & $45.12\pm 0.01$ & $1.26\pm 0.11$ & 2435.4 & 8.89 \\
DES J024831.08$+$005025.60 & 0.89 & $286^{+36}_{-102}$ & $341^{+103}_{-85}$ & 0.91 & 0.0 & $3.52\pm 0.15$ & $44.89\pm 0.02$ & $1.78\pm 0.23$ & 2130.3 & 8.76 \\
DES J024723.54$+$002536.50 & 1.86 & $867^{+64}_{-66}$ & $880^{+44}_{-30}$ & 0.84 & 0.0 & $3.73\pm 0.10$ & $45.89\pm 0.01$ & $3.68\pm 0.28$ & 2452.6 & 9.19 \\
DES J024944.09$+$003317.50 & 1.48 & $412^{+31}_{-67}$ & $402^{+66}_{-39}$ & 0.88 & 10.0 & $3.10\pm 0.06$ & $45.50\pm 0.01$ & $1.48\pm 0.12$ & 2207.4 & 8.83 \\
DES J024455.45$-$011500.40 & 1.53 & $165^{+28}_{-21}$ & $210^{+89}_{-39}$ & 0.77 & 8.0 & $1.47\pm 0.05$ & $45.22\pm 0.02$ & $1.93\pm 0.29$ & 3731.1 & 8.88 \\
DES J025225.52$+$003405.90 & 1.62 & $520^{+31}_{-26}$ & $505^{+43}_{-39}$ & 0.69 & 0.0 & $2.86\pm 0.06$ & $45.59\pm 0.01$ & $1.93\pm 0.18$ & 2637.9 & 9.06 \\
DES J022716.52$-$050008.30 & 1.64 & $524^{+14}_{-14}$ & $495^{+52}_{-49}$ & 0.68 & 2.1 & $1.94\pm 0.06$ & $45.44\pm 0.01$ & $1.45\pm 0.15$ & 2128.2 & 8.88 \\
DES J022751.50$-$044252.70 & 1.79 & $538^{+28}_{-13}$ & $558^{+33}_{-33}$ & 0.70 & 0.0 & $1.38\pm 0.05$ & $45.41\pm 0.02$ & $1.75\pm 0.18$ & 2849.7 & 9.12 \\
DES J022208.15$-$065550.50 & 1.66 & $439^{+25}_{-31}$ & $423^{+37}_{-40}$ & 0.96 & 11.5 & $1.09\pm 0.03$ & $45.20\pm 0.01$ & $1.90\pm 0.38$ & 2167.8 & 8.81 \\
DES J033836.19$-$295113.50 & 1.15 & $225^{+46}_{-51}$ & $280^{+85}_{-49}$ & 0.96 & 0.0 & $0.78\pm 0.04$ & $44.57\pm 0.02$ & $2.55\pm 0.38$ & 2987.0 & 8.90 \\
DES J033903.66$-$293326.50 & 1.68 & $308^{+40}_{-66}$ & $284^{+86}_{-30}$ & 0.67 & 10.0 & $0.67\pm 0.02$ & $45.01\pm 0.01$ & $1.64\pm 0.15$ & 3229.6 & 9.00 \\
DES J033328.93$-$275641.21 & 0.84 & $175^{+16}_{-17}$ & $182^{+35}_{-69}$ & 0.82 & 2.3 & $2.70\pm 0.22$ & $44.70\pm 0.04$ & $1.59\pm 0.16$ & 2549.0 & 8.72 \\
DES J022436.64$-$063255.90 & 1.42 & $181^{+25}_{-26}$ & $147^{+48}_{-44}$ & 0.83 & 0.0 & $0.83\pm 0.03$ & $44.88\pm 0.02$ & $1.36\pm 0.18$ & 3308.3 & 8.84 \\
DES J033211.42$-$284323.99 & 1.24 & $132^{+27}_{-21}$ & $137^{+69}_{-78}$ & 1.00 & 14.3 & $0.32\pm 0.03$ & $44.28\pm 0.04$ & $1.13\pm 0.19$ & 3388.9 & 8.75 \\
DES J033213.36$-$283620.99 & 1.49 & $171^{+13}_{-11}$ & $192^{+31}_{-45}$ & 0.85 & 3.6 & $2.23\pm 0.04$ & $45.37\pm 0.01$ & $1.23\pm 0.10$ & 2775.1 & 8.65 \\
DES J003710.86$-$444048.11 & 1.07 & $382^{+34}_{-30}$ & $430^{+29}_{-35}$ & 0.93 & 12.0 & $12.18\pm 0.20$ & $45.67\pm 0.01$ & $1.79\pm 0.04$ & 2398.1 & 8.95 \\
DES J003922.97$-$430230.41 & 1.37 & $564^{+53}_{-34}$ & $637^{+17}_{-41}$ & 0.76 & 17.2 & $2.80\pm 0.07$ & $45.35\pm 0.01$ & $1.32\pm 0.09$ & 2416.8 & 9.10 \\
DES J002933.85$-$435240.69 & 1.00 & $571^{+13}_{-29}$ & $540^{+32}_{-32}$ & 0.73 & 18.2 & $23.20\pm 0.26$ & $45.86\pm 0.00$ & $1.19\pm 0.07$ & 2257.1 & 9.09 \\
DES J003207.44$-$433049.00 & 1.53 & $376^{+10}_{-5}$ & $362^{+21}_{-34}$ & 0.65 & 3.0 & $3.40\pm 0.03$ & $45.59\pm 0.00$ & $2.06\pm 0.07$ & 1901.2 & 8.65 \\
DES J003015.00$-$430333.52 & 1.65 & $508^{+30}_{-25}$ & $467^{+28}_{-19}$ & 0.73 & 2.6 & $2.40\pm 0.04$ & $45.54\pm 0.01$ & $1.16\pm 0.04$ & 3915.7 & 9.39 \\
DES J003052.76$-$430301.08 & 1.43 & $383^{+35}_{-14}$ & $388^{+35}_{-37}$ & 0.78 & 5.3 & $3.46\pm 0.04$ & $45.50\pm 0.00$ & $1.93\pm 0.08$ & 2117.0 & 8.77 \\
DES J003232.61$-$433302.99 & 1.49 & $537^{+11}_{-14}$ & $535^{+43}_{-57}$ & 0.98 & 0.0 & $1.95\pm 0.05$ & $45.31\pm 0.01$ & $2.03\pm 0.08$ & 3927.3 & 9.45 \\
DES J003234.33$-$431937.81 & 1.64 & $656^{+63}_{-31}$ & $649^{+13}_{-16}$ & 0.71 & 0.0 & $3.36\pm 0.04$ & $45.67\pm 0.01$ & $1.72\pm 0.09$ & 1784.6 & 8.82 \\
DES J003206.50$-$425325.22 & 1.75 & $479^{+40}_{-26}$ & $466^{+26}_{-26}$ & 0.95 & 0.0 & $2.16\pm 0.06$ & $45.57\pm 0.01$ & $1.49\pm 0.07$ & 3769.7 & 9.32 \\
\hline
\end{tabular}
\caption{Characteristics of the 25 AGN in our final sample with successful Mg II lag measurements. Columns (1) and (2) give the DES name and redshift of the source. Columns (3) and (4) give the lags and uncertainties in the observer frame from JAVELIN and the ICCF method, respectively. Column (5) gives the probability within the major peak of the JAVELIN lag distribution (see Section \ref{subsec:cut_Y21}). Column (6) gives the false positive rate from simulations (see Section \ref{subsec:simulations}). Column (7) gives the spectral flux density and its uncertainty at 3000 \AA\, in unit of $10^{-17}$ erg s$^{-1}$ cm$^{-2}$ \AA$^{-1}$\,. Columns (8) gives the 3000 \AA\, monochromatic luminosity and its uncertainty. Column (9) gives the ratio $\mathcal{R}_{\rm Fe}$ of the iron flux to the Mg II flux (see Section \ref{subsec:RFe}). Columns (10) and (11) give the line dispersion and black hole mass. The black hole mass uncertainty is about 0.4 dex.}
\label{tab:result}
\end{table*}

\begin{table*}
\small
\renewcommand{\arraystretch}{1.25}
\begin{tabular}{cllll}
\hline
{ Source Name} & {$F_{\rm var}$ (g-band)} & {$F_{\rm var}$ (Mg II)} &  { $\chi^2_r$ (g-band)} & { $\chi^2_r$ (Mg II)} \\
 &   &   &   &  \\
\hline
DES J024340.09$+$001749.40 & $0.252\pm 0.001$ & $0.121\pm 0.013$ & 164650.0 & 6.6 \\
DES J025254.18$+$001119.70 & $0.141\pm 0.002$ & $0.117\pm 0.041$ & 128.8 & 3.0 \\
DES J024831.08$+$005025.60 & $0.196\pm 0.002$ & $0.186\pm 0.025$ & 240.2 & 4.8 \\
DES J024723.54$+$002536.50 & $0.175\pm 0.001$ & $0.228\pm 0.029$ & 577.2 & 4.2 \\
DES J024944.09$+$003317.50 & $0.113\pm 0.001$ & $0.120\pm 0.022$ & 3308.1 & 4.2 \\
DES J024455.45$-$011500.40 & $0.205\pm 0.002$ & $0.207\pm 0.017$ & 485.3 & 7.0 \\
DES J025225.52$+$003405.90 & $0.120\pm 0.002$ & $0.111\pm 0.019$ & 46.4 & 3.4 \\
DES J022716.52$-$050008.30 & $0.217\pm 0.001$ & $0.114\pm 0.012$ & 3477.1 & 5.0 \\
DES J022751.50$-$044252.70 & $0.199\pm 0.001$ & $0.244\pm 0.016$ & 23298.1 & 7.5 \\
DES J022208.15$-$065550.50 & $0.128\pm 0.002$ & $0.213\pm 0.044$ & 62.5 & 4.4 \\
DES J033836.19$-$295113.50 & $0.222\pm 0.005$ & $0.296\pm 0.023$ & 54.6 & 8.3 \\
DES J033903.66$-$293326.50 & $0.161\pm 0.004$ & $0.185\pm 0.037$ & 35.8 & 2.7 \\
DES J033328.93$-$275641.21 & $0.305\pm 0.003$ & $0.140\pm 0.016$ & 281561.2 & 7.0 \\
DES J022436.64$-$063255.90 & $0.263\pm 0.003$ & $0.098\pm 0.026$ & 321.5 & 3.8 \\
DES J033211.42$-$284323.99 & $0.311\pm 0.006$ & $0.244\pm 0.023$ & 94.9 & 8.9 \\
DES J033213.36$-$283620.99 & $0.177\pm 0.001$ & $0.163\pm 0.015$ & 5525.8 & 6.1 \\
DES J003710.86$-$444048.11 & $0.084\pm 0.001$ & $0.084\pm 0.009$ & 158385.6 & 5.2 \\
DES J003922.97$-$430230.41 & $0.160\pm 0.001$ & $0.140\pm 0.019$ & 1995.1 & 3.2 \\
DES J002933.85$-$435240.69 & $0.073\pm 0.001$ & $0.174\pm 0.020$ & 2608.0 & 3.9 \\
DES J003207.44$-$433049.00 & $0.096\pm 0.001$ & $0.102\pm 0.015$ & 127.4 & 3.4 \\
DES J003015.00$-$430333.52 & $0.078\pm 0.002$ & $0.046\pm 0.005$ & 219.0 & 4.3 \\
DES J003052.76$-$430301.08 & $0.085\pm 0.001$ & $0.049\pm 0.008$ & 662.2 & 2.8 \\
DES J003232.61$-$433302.99 & $0.251\pm 0.002$ & $0.086\pm 0.008$ & 405.7 & 4.7 \\
DES J003234.33$-$431937.81 & $0.070\pm 0.002$ & $0.075\pm 0.018$ & 63.4 & 1.8 \\
DES J003206.50$-$425325.22 & $0.172\pm 0.001$ & $0.206\pm 0.013$ & 243.3 & 6.8 \\
\hline
\end{tabular}
\caption{Lightcurve variability of the final sample. Column (1) gives DES name of the source. Columns (2) and (3) give the fractional variability $F_{\rm var}$ defined by Equation (\ref{eq:Fvar}) of the g-band and Mg II lightcurves, respectively. Columns (4) and (5) give the $\chi^2_r$ value defined by Equation (\ref{eq:chi2_lcvar}).}
\label{tab:lcvar}
\end{table*}

\begin{figure*}
\includegraphics[width=1.\linewidth]{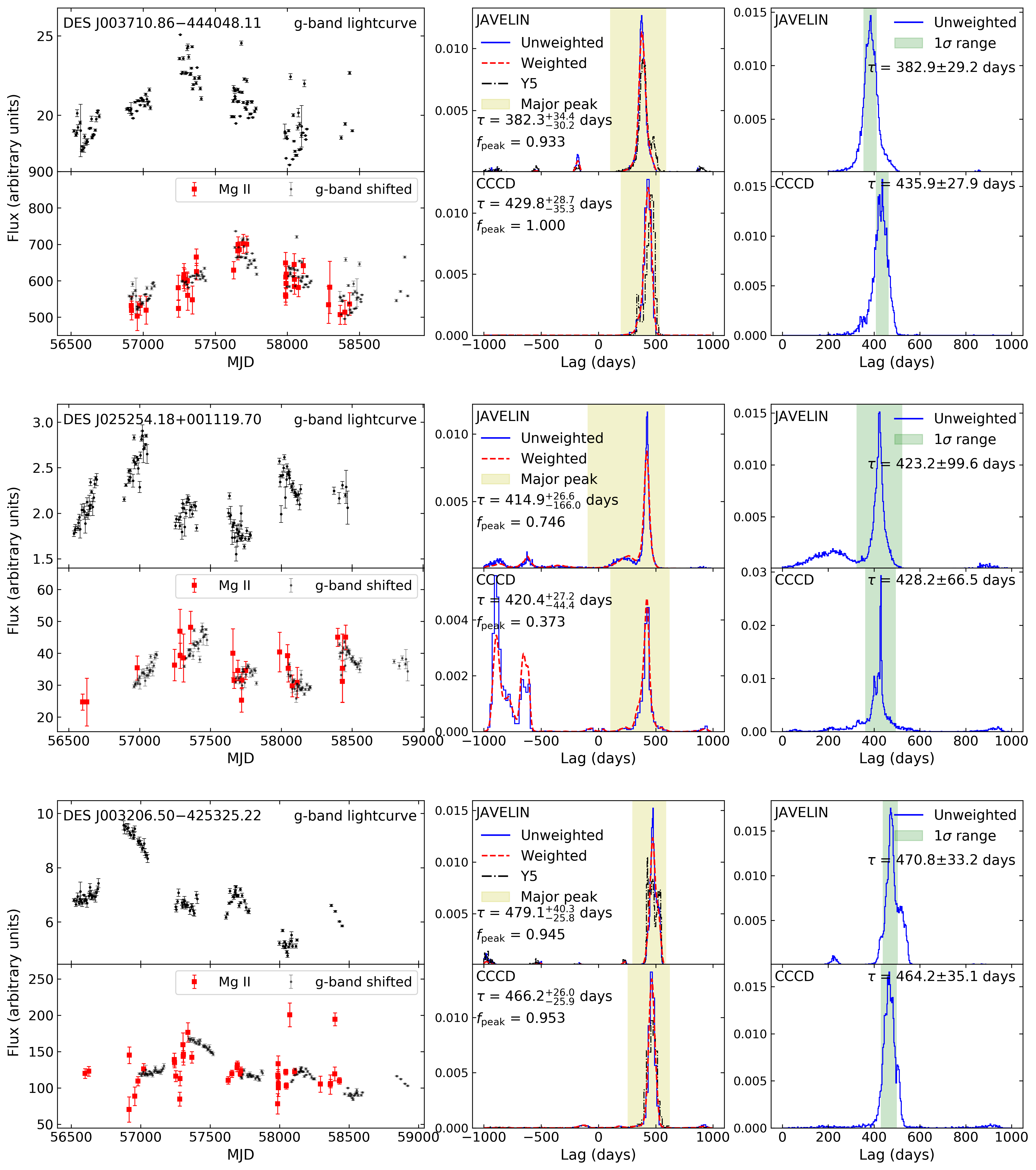}
\caption{Lightcurves and lag measurements. Each of the three main rows is for an AGN with its name given in the upper left corner. ({\it Left column}) The upper and lower panels show the $g$-band lightcurve and the Mg II line lightcurve, respectively. The black points in the lower panel show the $g$-band lightcurve shifted by the best-fit lag. ({\it Middle column}) The upper and lower panels show the lag distributions from JAVELIN and ICCF with a symmetric lag prior. The blue solid and red dashed lines represent the unweighted and weighted lag distributions, respectively. For sources presented in Y21, the black dash-dotted lines show the lag distributions from the 5-yr lightcurves. The yellow shaded area marks the major peak region. ({\it Right column}) Lag distributions with a positive lag prior. The green shaded area represents the $1\sigma$ range centered at the peak, where $\sigma$ is defined as the mean absolute deviation of the lag distribution.}
\label{fig:lag_examp}
\end{figure*}

\subsection{Symmetric Prior} \label{subsec:cut_Y21}
Method 1 resembles the lag analysis in Y21 and the SDSS RM project \citep[e.g.,][]{Grier2019,Homayouni2020}. We set a lag prior range [$\tau_{\rm min}=-1000$ days, $\tau_{\rm max}=1000$ days] for both JAVELIN and ICCF and define a weighting function to suppress aliasing. The weighting function is the convolution of two components. The first component is defined as 
\begin{equation}
P(\tau) = [N(\tau)/N(0)]^2
\label{eq:weight_sdss}
\end{equation}
where $N(\tau)$ is the number of overlapping points between the continuum lightcurve shifted by the time lag $\tau$ and the line lightcurve. This penalizes lags in the seasonal gaps where the shifted continuum has little overlap with the line lightcurve. The second component is the auto-correlation function (ACF) of the continuum lightcurve, which characterizes how fast the continuum varies. We set ${\rm ACF}=0$ when it is below zero. 

We multiply the lag distribution by the weighting function and convolve the weighted distribution with a Gaussian kernel that has a width of 12 days, the same width used in Y21 and \citet{Homayouni2020}. We define the major peak as the highest peak in the weighted and smoothed lag distribution. When there are multiple connected peaks, we define one as a separate peak if its prominence exceeds 10\% of the prominence of the neighboring peaks and it is separated from the neighboring peaks by at least 10 days. We use the package \textsf{scipy.signal.peak\_prominences} \citep{scipy} to calculate the peak prominence defined as the vertical distance between the peak and the higher minimum at its two sides. We then define the lag as the median of the unweighted lag distribution within the major peak and define the lag uncertainty based on its 16th and 84th percentiles. 

We define the first set of selection criteria based on Method 1:
\begin{enumerate}[label=(\alph*)]
\item $f_{\rm peak} > 0.6$ for the JAVELIN lag distribution, where $f_{\rm peak}$ is the probability within the major peak;  
\item The lag uncertainty from JAVELIN is less than 110 days; and
\item The lags from JAVELIN and ICCF agree within $2\sigma$.
\end{enumerate}

The first criterion ensures that the major peak has enough significance to be distinguished from the aliasing peaks. The second criterion requires that the lag distribution has enough constraining power and excludes lag distributions that are flat over a wide range of lags. The agreement between JAVELIN and ICCF required by the third criterion helps to increase the lag reliability. We obtain 62 lags that pass the Method 1 criteria, 48 of which are positive.

\subsection{Positive Prior} \label{subsec:cut_P21}
Method 2 is defined by \citet{Penton2022} and is based on simulated lightcurves that have the same cadence and SNR as the OzDES observations. We use a lag prior range [$\tau_{\rm min}=0$ day, $\tau_{\rm max}=1000$ days] to derive the lag distribution from JAVELIN and ICCF. We analyze the full lag distribution binned by 3 days without any weighting or major peak identification. We define the second set of selection criteria based on Method 2:
\begin{enumerate}[label=(\alph*)]
\item The mean absolute deviation of the JAVELIN lag distribution is less than 110 days;  
\item The separation between the median and peak of the JAVELIN lag distribution is less than 110 days; and
\item The separation between the peaks of the JAVELIN and ICCF lag distribution is less than 100 days.
\end{enumerate}

The threshold values in each criterion are the same as \citet{Penton2022}. The first and second criteria effectively require the lag distribution to have a strong major peak compared to the aliasing peaks, since the aliasing peaks would increase the mean absolute deviation and result in a difference between the median and peak. There are 56 quasars that pass the Method 2 criteria.

\subsection{Final Measurements} \label{subsec:cut_main}
We define a successful lag measurement as one which passes both the Method 1 and 2 criteria. This yields 25 lag measurements which we refer to as our final sample. We adopt the lag and uncertainty estimates in Section \ref{subsec:cut_Y21} based on the JAVELIN lag distributions, since previous studies showed that JAVELIN provides better lag and uncertainty estimates than the ICCF method \citep[e.g.,][]{Li2019,Yu2020_RMErr}. Table \ref{tab:result} gives the lag measurements of the final sample. The final sample spans a redshift range of $0.84 \leq z \leq 1.86$ and an observed lag range of $130\,{\rm days} < \tau < 880\,{\rm days}$.

Table \ref{tab:lcvar} gives two metrics of the lightcurve variability of the final sample. The first metric is the fractional variability $F_{\rm var}$ \citep[e.g.,][]{Rodriguez1997,Vaughan2003} defined as
\begin{subequations}
\begin{align}
& F_{\rm var} = \frac{\sqrt{S^2 - \langle \sigma^2 \rangle}}{\langle F \rangle} \\
& \sigma_{F_{\rm var}} = \sqrt{\left(\sqrt{\frac{1}{2N}} \frac{\langle \sigma^2 \rangle}{F_{\rm var} \langle F \rangle^2}\right)^2 + \left(\sqrt{\frac{\langle \sigma^2 \rangle}{N}} \frac{1}{\langle F \rangle}\right)^2}
\end{align}
\label{eq:Fvar}
\end{subequations}
where $S$ is the standard deviation of the lightcurve, $\langle \sigma^2 \rangle$ is the mean square of the flux uncertainties, $\langle F \rangle$ is the mean flux, $\sigma_{F_{\rm var}}$ is the uncertainty of $F_{\rm var}$, and $N$ is the number of epochs. The second metric $\chi^2_r$ is defined as 
\begin{subequations}
\begin{align}
& \chi^2 = \sum \limits_i  \frac{(F_{i}-\langle F \rangle)^2}{\sigma_{i}^2} \\
& \chi^2_r = \chi^2 / (N-1)
\end{align}
\label{eq:chi2_lcvar}%
\end{subequations}
where $F_{i}$ and $\sigma_{i}$ are the flux and uncertainty of the $i^{\rm th}$ epoch. The fractional variability $F_{\rm var}$ characterizes the excess variability amplitude relative to the mean flux, while $\chi^2_r$ characterizes the significance of the variability relative to the uncertainties.

Figure \ref{fig:lag_examp} shows examples of the lightcurves and lag distributions of our final sample. The lag distributions exhibit clear major peaks. The major peaks are significantly stronger than the secondary peaks for the JAVELIN lag distributions. The median JAVELIN major peak fraction $f_{\rm peak}$ of the final sample is 0.78. The lags from the symmetric and positive prior ranges all agree within $1\sigma$. The JAVELIN and ICCF lags are all consistent within 100 days and within $2\sigma$, as required by the selection criteria. 

The lightcurves in Figure \ref{fig:lag_examp} exhibit clear variability features. The black points in the lower left panels show the continuum lightcurve shifted by the best-fit JAVELIN lag with the symmetric prior range. For the sources shown in the top two rows, the shifted continuum lightcurve matches the line lightcurve well, which supports the reliability of the lag measurement. The lag of the third source is close to the 1.5-yr seasonal gap, but its shifted continuum lightcurve is a reasonable ``interpolation'' between the line lightcurves in different seasons. 

A special feature of DES J003206.50$-$425325.22 (bottom panel of Figure \ref{fig:lag_examp}) is the two outliers at MJD 58073 and 58399. The excess flux is due to a significant broadening of the Mg II line. This can be intrinsic to the Mg II line or due to a drastic change in the iron emission that is not characterized by the model. The JAVELIN lag distribution would be dominated by a sharp aliasing peak at $\sim 540$ days if we included these two epochs, which differs from the $\sim 470$-day ICCF lag. For lags of a few hundred days, we expect the signal to come from lightcurve features over several years rather than dramatic short time scale variability, so we exclude these two epochs for this particular source. Excluding these two epochs also gives a cleaner JAVELIN lag distribution compared to the Y5 results where we kept the MJD 58073 epoch, and the best-fit lag changes by $\sim 40$ days relative to Y5. 

Seven of the nine quasars from Y21 pass our final sample criteria in this paper after adding the Y6 data. The black dashed lines in the top and bottom panels of Figure \ref{fig:lag_examp} show the lag distributions presented in Y21 based on the first 5-yr lightcurves. The new and old lags agree within $1\sigma$ except for the case of DES J003206.50$-$425325.22 discussed above. Adding the Y6 data suppresses some aliasing peaks and makes the lag signal cleaner. Two quasars from Y21 (DES J021612.83$-$044634.10 and DES J033553.51$-$275044.70) fail to pass the lag selection. While both pass the Method 1 criteria, they are excluded by the Method 2 criteria due to large mean absolute deviations and differences between the JAVELIN and ICCF lags caused by aliasing peaks.

\section{Lag Reliability} \label{sec:reliability}
We assess the reliability of lag measurements using simulations based on both the DRW model and the observed lightcurves of intensively monitored local AGN. We then discuss the lag reliability of our final sample and compare with the samples selected by only one of the criteria. 

\subsection{Simulations with DRW} \label{subsec:simulations}
We adopt the simulation tool developed by \citet{Penton2022} to create simulated lightcurves following the same procedure as Y21. For each of the 25 final sample sources that pass both sets of criteria for Method 1 and 2, we create 1000 realizations of DRW lightcurves that have the same variability as the observed continuum lightcurve. We use the DRW lightcurve as the simulated continuum and convolve it with a top-hat transfer function to create the simulated line lightcurves. The input lag $\tau_{\rm i}$ of the transfer function is randomly drawn from a uniform distribution between 10 days and 1000 days. We then re-sample and add noise to the simulated lightcurves so that their cadence and SNR match the observed lightcurves. 

\begin{figure}
\includegraphics[width=\linewidth]{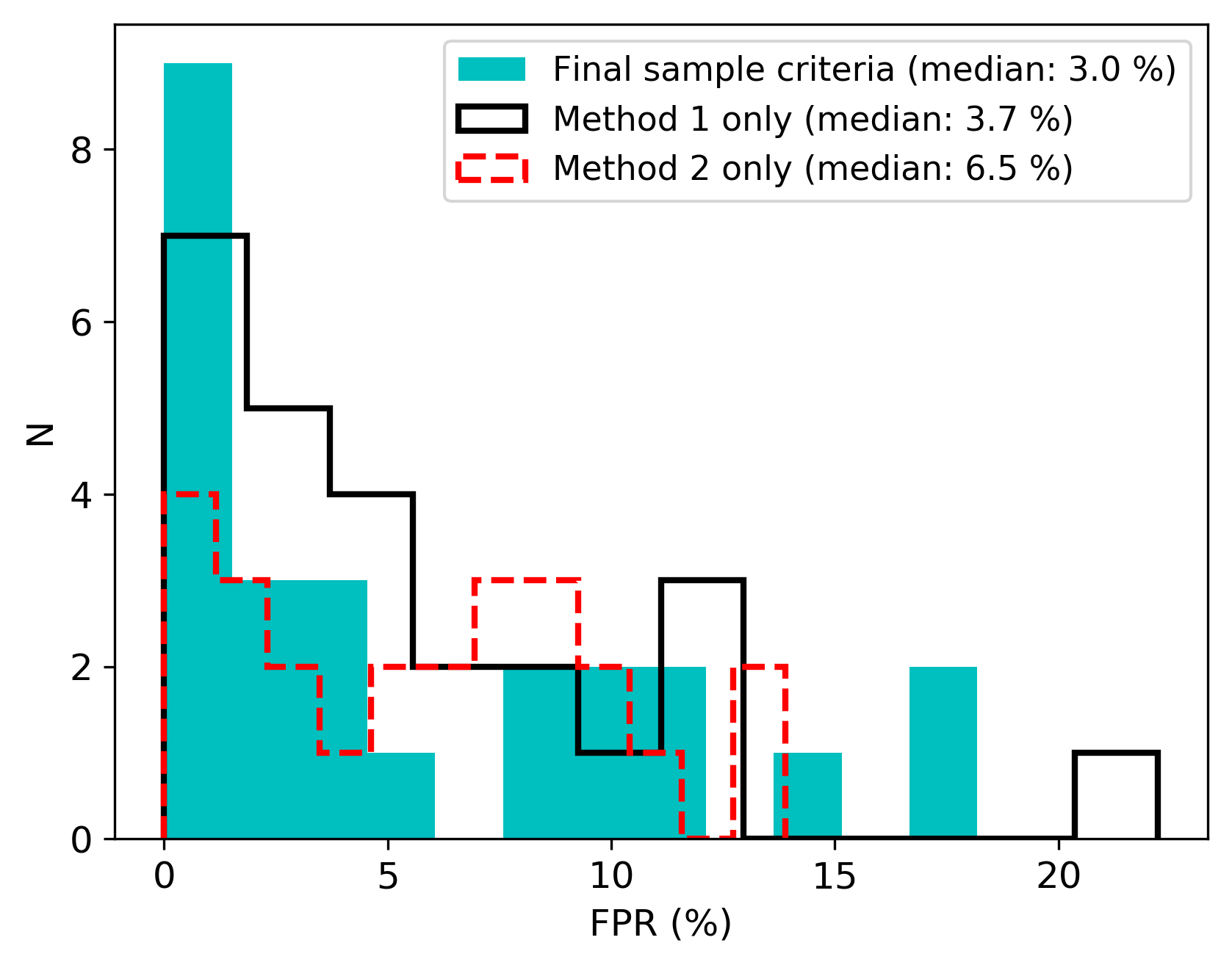}
\caption{False positive rate distribution of the final sample. The cyan filled histograms are from simulations where the lag quality criteria are same as the final sample (Method 1 + 2). The black solid and red dashed lines are from simulations where we only use the Method 1 criteria or the Method 2 criteria to select successful lag measurements, respectively.}
\label{fig:FPR}
\end{figure}

We then use the method described in Section \ref{sec:lcanl} to measure lags from the simulated lightcurve pairs. We define a lag as a ``false positive'' if the measured lag $\tau_{\rm m}$ differs from the the input lag $\tau_{\rm i}$ by $3\sigma$. For a final sample source with an observed lag $\tau_{\rm obs}$, we define the false positive rate (FPR) as 
\begin{equation}
{\rm FPR} = N_{\rm bad,m} (\tau_{\rm obs}) / N_{\rm p,m} (\tau_{\rm obs})
\label{eq:FPR_obs}
\end{equation}
where $N_{\rm p,m} (\tau_{\rm obs})$ is the number of realizations where the measured lag $\tau_{\rm m}$ passes the final sample selection and is within $1\sigma$ of $\tau_{\rm obs}$ , and $N_{\rm bad,m} (\tau_{\rm obs})$ is the number of false positives among the $N_{\rm p,m} (\tau_{\rm obs})$ realizations. The FPR is an estimate of the probability that the observed lag $\tau_{\rm obs}$ is different from the true lag. Table \ref{tab:result} gives the FPR of each AGN in the final sample. The filled histograms in Figure \ref{fig:FPR} show the FPR distribution of the final sample. The median FPR is $\sim 3\%$. The sum of the FPR for all 25 AGN is $\sim 1.3$, which indicates that there could be one incorrect lag in the sample. The median and average of the FPR are similar to the Y21 sample where the median was $\sim 4\%$ and the sum was $0.44$ for 9 AGN. 

\subsection{Simulations with Reference Lightcurves} \label{subsec:simreflc}
\begin{table}
\small
\renewcommand{\arraystretch}{1.25}
\begin{tabular}{cccc}
\hline
Name & Continuum Band & Line & Source \\
 &  &  &  \\
\hline
NGC 5548 & 1367 \AA & Ly $\alpha$ & \citet{DeRosa2015} \\
Mrk 335 & V band & H$\beta$ & \citet{Grier2012} \\
Mrk 1501 & V band & H$\beta$ & \citet{Grier2012} \\
3C 120 & V band & H$\beta$ & \citet{Grier2012} \\
Mrk 6 & V band & H$\beta$ & \citet{Grier2012} \\
PG 2130+099 & V band & H$\beta$ & \citet{Grier2012} \\
Mrk 704 & 5100 \AA & H$\beta$ & \citet{DeRosa2018} \\
NGC 3227 & 5100 \AA & H$\beta$ & \citet{DeRosa2018} \\
NGC 3516 & 5100 \AA & H$\beta$ & \citet{DeRosa2018} \\
NGC 4151 & 5100 \AA & H$\beta$ & \citet{DeRosa2018} \\
NGC 5548 & 5100 \AA & H$\beta$ & \citet{DeRosa2018} \\
Mrk 142 & 5100 \AA & H$\beta$ & \citet{Du2014} \\
Mrk 382 & V band & H$\beta$ & \citet{Wang2014} \\
\hline
\end{tabular}
\caption{The AGN used for the simulations in Section \ref{subsec:simreflc}. Column (1) gives the object name. These objects generally have lightcurves for multiple continuum bands and emission lines. Columns (2) and (3) give the continuum band and emission line we adopt for the reference lightcurves. Column (4) gives of the source of the reference lightcurves. NGC 5548 has lightcurves from different seasons which we use as separate reference lightcurves.}
\label{tab:simreflc}
\end{table}

\begin{figure*}
\includegraphics[width=1.\linewidth]{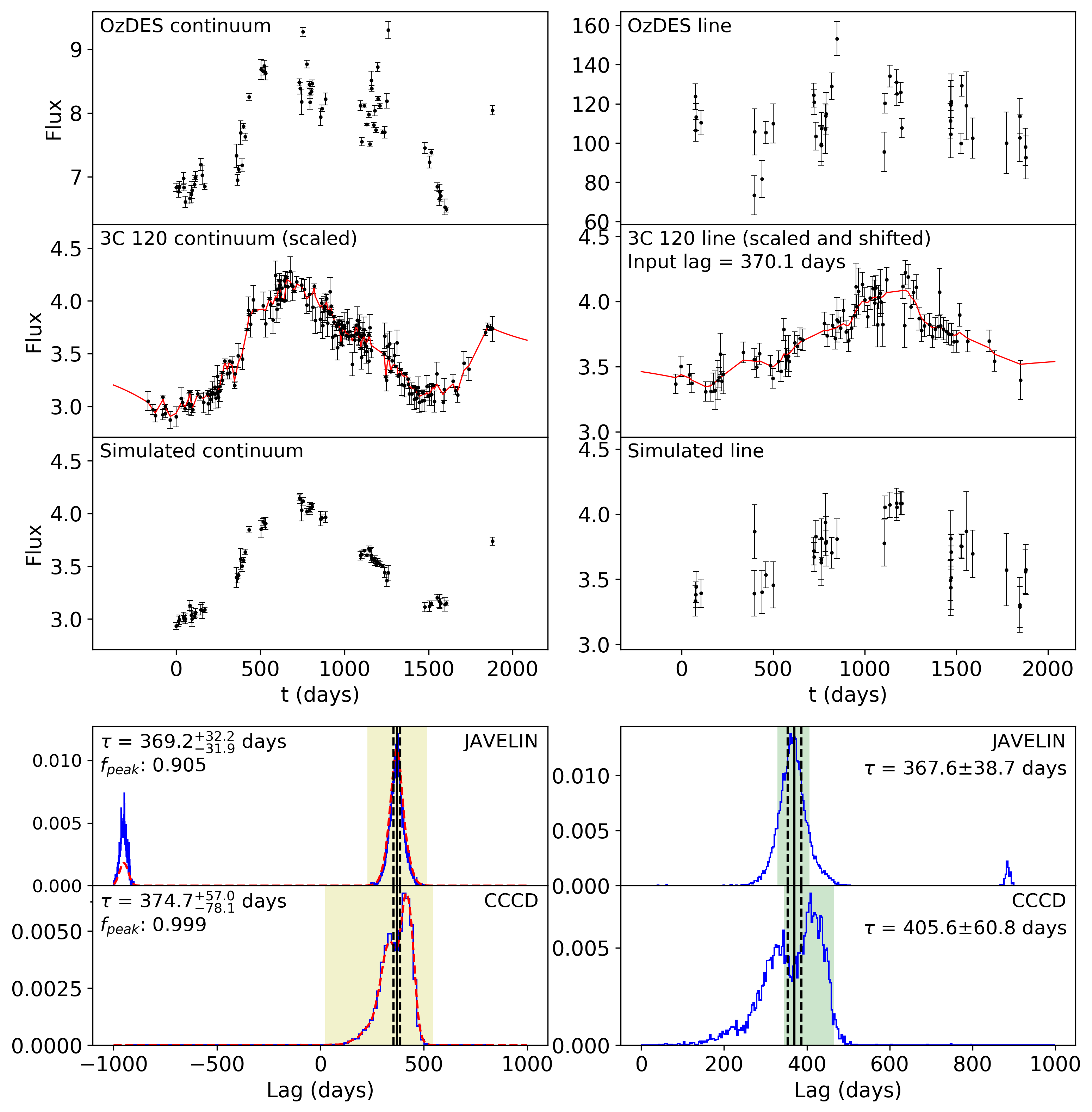}
\caption{Examples of the simulations based on the observed lightcurves of local AGN. ({\it Upper 3$\times$2 panels}) Procedure for creating simulated lightcurves based on the reference lightcurves. The lightcurves are in arbitrary units and the absolute flux scales have no effect on our simulations. The left and right columns show the continuum and line lightcurves, respectively. The top row shows the OzDES continuum and line lightcurves of DES J003207.44$-$433049.00. The middle row shows the scaled reference lightcurves based on 3C 120. The input lag between the scaled reference lightcurves is 370.1 days, which equals the observed lag between the OzDES lightcurves of DES J003207.44$-$433049.00. The red lines represent the best-fit lightcurves from JAVELIN. The bottom row shows the simulated lightcurves, which have the same cadence and variability amplitude as the OzDES lightcurves in the top row. ({\it Lower 2$\times$2 panels}) Lag distributions from the simulated lightcurves. The left and right columns show results from the symmetric and positive lag priors, respectively. The black solid line is drawn at the input lag. The black dashed lines show the $1\sigma$ uncertainty of the input lag, which is the uncertainty of the lag between the original reference lightcurves multiplied by the baseline scaling factor $X_t$. Other symbols have the same meanings as Figure \ref{fig:lag_examp}.}
\label{fig:simreflc_examp}
\end{figure*}

The simulations in Section \ref{subsec:simulations} assume that the AGN variability is a DRW stochastic process and that the line lightcurve is related to the continuum lightcurve by a top-hat transfer function. These assumptions may not hold in real AGN. For example, the AGN variability could deviate from the DRW model on time scales below $\sim$ month \citep[e.g.,][]{Mushotzky2011,Kasliwal2015,Smith2018}. AGN could also have ``BLR holidays'' when the broad line variability is not correlated with the continuum variability \citep[e.g.,][]{Goad2016,Horne2020}, and this could hamper the lag recovery \citep{Yu2020_RMErr}. 

We perform additional simulations using the observed lightcurves of several intensively monitored local AGN as ``reference lightcurves''. We first scale the time axis of the reference lightcurve by a factor of $X_t$ so that its baseline length matches the 6-yr baseline of the OzDES continuum and line lightcurves. While scaling the baseline increases the variability time scale, \citet{Stone2022} found that the DRW time scale $\tau_{\rm drw}$ of AGN could span $\sim 1.5$ orders of magnitude, so the scaling does not make the variability time scale unrealistic for AGN. We then shift the scaled reference lightcurve of the emission line by $\Delta t = \tau_{\rm i} - X_t \tau_{\rm ref}$ so that the lag between the shifted, scaled reference lightcurves of the continuum and emission line equal the desired input lag $\tau_{\rm i}$, where $\tau_{\rm ref}$ is the lag between original reference lightcurves from literature. The second row of Figure \ref{fig:simreflc_examp} shows an example of the scaled reference lightcurves. 

We interpolate the scaled reference continuum and line lightcurves using the predicted lightcurves from JAVELIN (red solid line in Figure \ref{fig:simreflc_examp}) and re-sample them to the same cadence as the OzDES lightcurves (top row of Figure \ref{fig:simreflc_examp}) to create the simulated lightcurves. We then assign an uncertainty $\sigma_{{\rm sim},j} = K \sigma_{{\rm oz},j}$ to each simulated epoch, where $j$ corresponds to the $j^{\rm th}$ epoch, $K$ is a constant coefficient and $\sigma_{{\rm oz},j}$ is the uncertainty of the corresponding OzDES epoch. The constant $K$ is derived such that the variability 
\begin{equation}
\chi_{\rm var}^2 = \sum_j \frac{(f_j - \langle f \rangle)^2}{\sigma_j^2}
\label{eq:lc_var}
\end{equation}
of the simulated lightcurve equals that of the OzDES lightcurve, where $f_j$ and $\sigma_j$ are the flux and uncertainty of the $j^{\rm th}$ epoch and $\langle f \rangle$ is the mean flux. If the assigned uncertainty $\sigma_{{\rm sim},j}$ is larger than the original uncertainty $\sigma_{0,j}$ of the reference lightcurve, we add additional Gaussian noise with a variance of $\sigma_{{\rm sim},j}^2 - \sigma_{0,j}^2$ to the simulated epoch. We do not add additional noise when $\sigma_{{\rm sim},j}<\sigma_{0,j}$, so some simulated lightcurve may have underestimated noise relative to the observed lightcurves. This will lead to an underestimate of the lag uncertainty from the simulated lightcurves and an overestimate of the false positive rate. It is not a significant problem for our purpose, since it puts an even higher requirement on the lag quality. While matching the overall variability of the lightcurves does not guarantee the short time scale variability is identical, we expect the major lag signal to be from lightcurve features over multiple years, so this approximation would not significantly affect our simulation results. The third row of Figure \ref{fig:simreflc_examp} shows an example of a final simulated lightcurve. 

Each reference AGN effectively gives a realization of the simulated lightcurve. We use 12 reference AGN summarized in Table \ref{tab:simreflc} from \citet{Grier2012}, \citet{Du2014}, \citet{Wang2014}, \citet{DeRosa2015} and \citet{DeRosa2018} where the cadences of the scaled reference lightcurves are higher than 15 days and 20 days for the continuum and emission line, respectively. All but Mrk 382 have velocity resolved RM results \citep{Grier2013Peterson,DeRosa2018,YLi2018,Horne2020}. NGC 5548 has lightcurves for two different seasons, which we use as separate reference lightcurves, so there are in total 13 realizations of simulated lightcurves. For each realization, we use two different sets of input lag, cadence and variability: $\tau_{\rm i} = 370.1$ days that mimics the OzDES lightcurves of DES J003207.44$-$433049.00 and $\tau_{\rm i} = 540.8$ days that mimics the OzDES lightcurves of DES J003232.61$-$433302.99. These input lags characterize two regimes of lag measurements where the shifted continuum overlaps well with the line lightcurve and where much of it falls in the seasonal gaps of the line lightcurve. The lower 2$\times$2 panels of Figure \ref{fig:simreflc_examp} shows an example of the lag distributions from the simulated lightcurves, and the recovered lags are consistent with the input lag. The same figures for other reference AGN and input lags are available in the supplementary material. There are three realizations for $\tau_{\rm i} = 370.1$ days and six realizations for $\tau_{\rm i} = 540.8$ days that pass the final sample criteria. None of them are ``false positives'' as defined in Section \ref{subsec:simulations}. While this simulation still does not perfectly mimic the behavior of the Mg II line due to the difference between H$\beta$ and Mg II and the simplifications used in the method, it is a qualitative verification of our lag selection criteria based on real AGN lightcurves rather than idealized models.

\subsection{Comparison of Lag Selection Criteria} \label{subsec:cricomp}
\begin{figure*}
\includegraphics[width=1.05\linewidth]{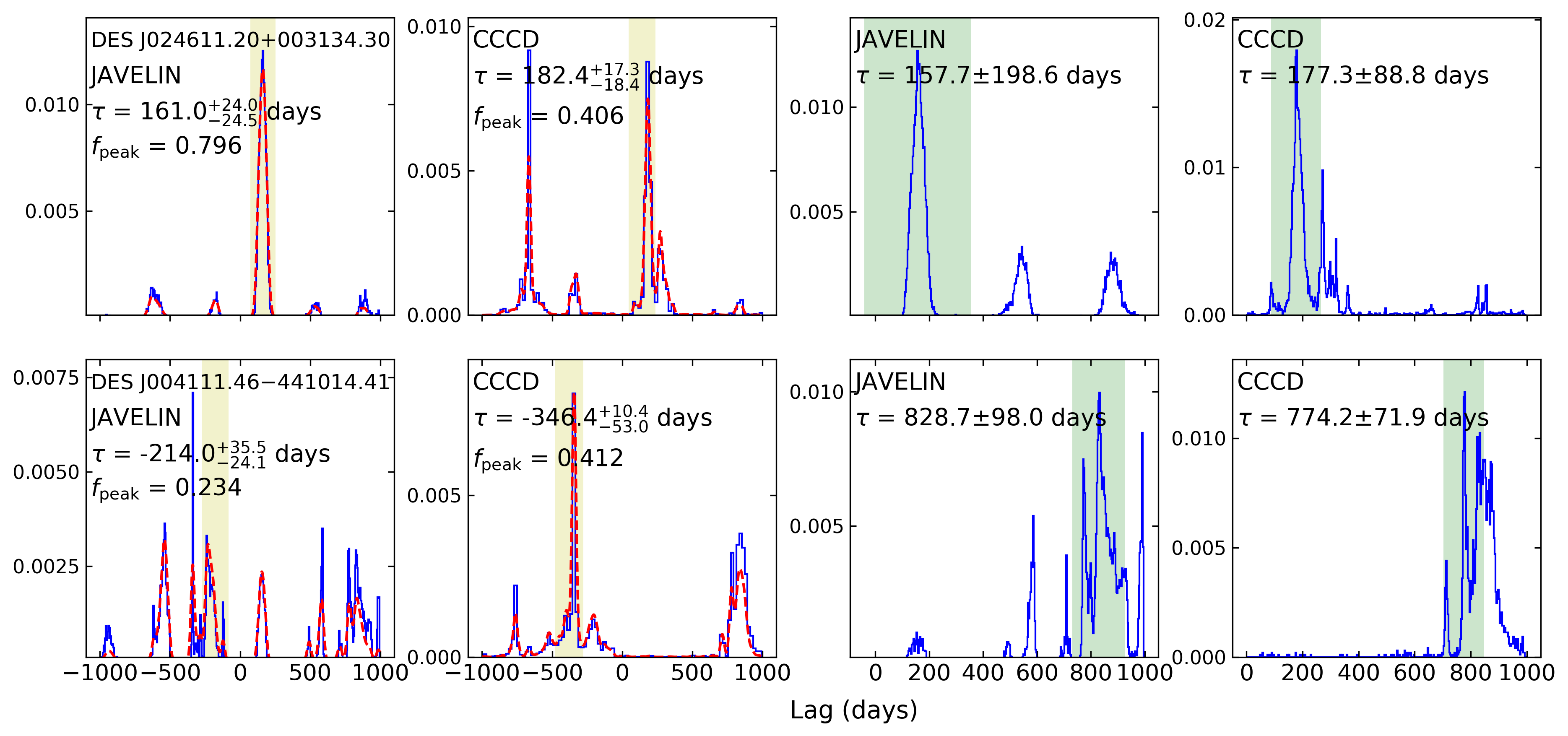}
\caption{Lag distributions of DES J024611.20+003134.30 (upper row) which only passes the Method 1 criteria and DES J004111.46$-$441014.41 (lower row) which only passes the Method 2 criteria. The two leftmost columns show the JAVELIN and ICCF lag distributions with a symmetric prior, while the two rightmost columns show results with a positive prior. Other symbols have the same meanings as Figure \ref{fig:lag_examp}.}
\label{fig:lag_onecut}
\end{figure*}

In addition to the final sample, there are AGNs that only pass the criteria for Method 1 or 2. Figure \ref{fig:lag_onecut} shows examples of the lag distributions for these AGNs. DES J024611.20+003134.30 passes the Method 1 criteria with a high major peak fraction $f_{\rm peak}$ of the JAVELIN lag distribution, while it is excluded by the Method 2 criteria due to the large mean absolute deviation caused by the aliasing peaks. DES J004111.46$-$441014.41 marginally passes the Method 2 criteria with a mean absolute deviation close to the threshold, while it is excluded by the Method 1 criteria due to significant aliasing when allowing negative lags. The lag distributions of both examples are ambiguous and dominated by aliasing signals at the seasonal gaps. This qualitatively shows how combining the two criteria can help exclude ambiguous measurements. The symmetric lag prior range in Method 1 reduces the strength of spurious signals caused by only allowing positive lags, while Method 2 is more sensitive to spurious lag distributions with multiple aliasing peaks at significantly different lags. 

To quantitatively compare the reliability of different samples, we repeat the FPR calculation in Section \ref{subsec:simulations} but use only one set of criteria to select successful lag measurements. Figure \ref{fig:FPR} compares the FPR from a single set of criteria to that from the final sample criteria. Using just one set of criteria results in a larger overall FPR than using both. We then analyze the simulated lightcurves from Section \ref{subsec:simreflc} using only one set of selection criteria. For an input lag $\tau_{\rm i} = 370.1$ days, we get three incorrect lag recoveries when only using the Method 1 criteria. For $\tau_{\rm i} = 540.8$ days, we get one incorrect lag recovery when only using the Method 2 criteria. This indicates that only using one set of criteria is less robust than the combined criteria. 

Another commonly used method of assessing the sample reliability is comparing the number of positive and negative lag measurements \citep[e.g.,][]{Grier2019,Homayouni2020,Yu2021}. The negative lags are from artifacts, so the comparison between the number of positive and negative lags gives an overall estimate of the contamination from spurious detection. For Method 1, we get 14 negative lags and 48 positive lags. The comparison is not directly applicable to Method 2 since it only allows positive lags when running JAVELIN and ICCF. We do a one-sided search of [$-1000$,$0$] days as symmetric to the positive prior [$0$,$1000$] days while keeping other criteria of Method 2 the same. This gives 29 negative lags compared to 56 positive lags. When combing the two sets of criteria, we get the 25 positive lags in the final sample and no negative lags. These comparisons indicate that the final sample is less contaminated by artifacts than the samples obtained using only a single set of criteria. There are likely to be physical lag measurements in the single-method samples. However, since they are more likely to be contaminated by spurious lags than our final sample and could increase the risk of biasing the $R-L$ relation, we do not include them in our $R-L$ relation. The lightcurves of the single-method sample are available in the supplementary material.

\section{Black Hole Mass and $R-L$ Relation} \label{sec:RL_BHMass}
\begin{figure*}
\includegraphics[width=1.\linewidth]{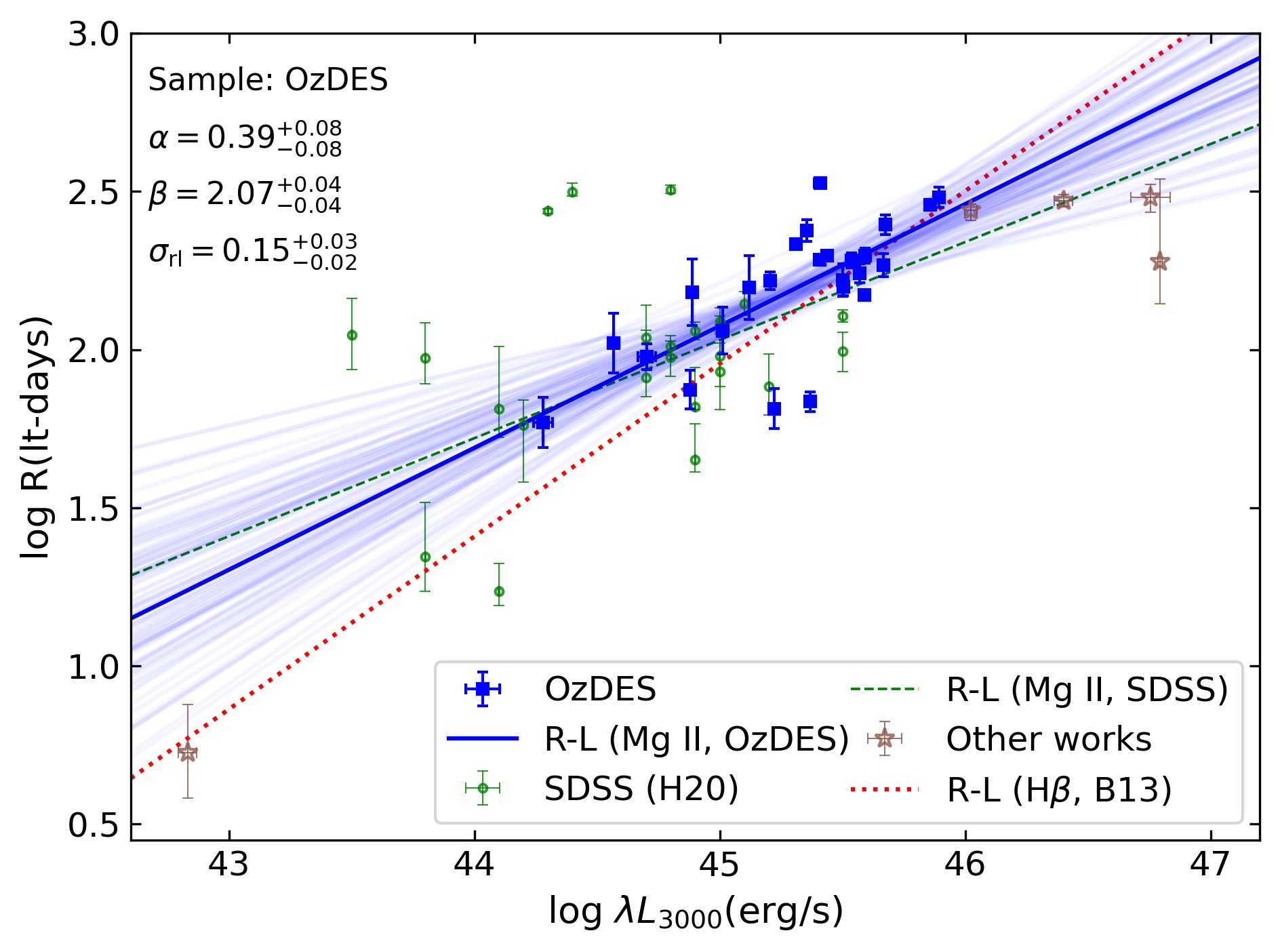}
\caption{Mg II $R-L$ relation from the OzDES RM project and previous works. The blue squares are our high-quality measurements from the full six years of data. The green circles are the gold sample of \citet{Homayouni2020}. The brown stars are measurements from \citet{Metzroth2006}, \citet{Lira2018}, \citet{Zajacek2020}, \citet{Zajacek2021} and \citet{Prince2022}. The blue solid line shows the best-fit Mg II $R-L$ relation constrained using our final sample with the slope $\alpha$ and intrinsic scatter $\sigma_{\rm rl}$ shown in the upper left corner. The lighter blue lines are 100 realizations randomly drawn from the MCMC chain, which illustrates the uncertainty of our $R-L$ relation. The green dashed line shows the Mg II $R-L$ relation from \citet{Homayouni2020}. The red dotted line shows the H$\beta$ $R-L$ relation from \citet{Bentz2013} with a slope of $\sim 0.5$ and an intrinsic scatter of $\sim 0.13$ dex. We use the bolometric correction from \citet{Richards2006} to convert the 3000 \AA\, monochromatic luminosity to 5100 \AA\, for the H$\beta$ $R-L$ relation.}
\label{fig:RL}
\end{figure*}

We calculate the black hole mass with Equation \ref{eq:bhmass}, and parameterize the line width using the line dispersion defined by 
\begin{equation}
\sigma_{\rm line}^2 = \left[\int \lambda^2 P(\lambda) d\lambda \, \Big/ \int P(\lambda) d\lambda \right] - P_0(\lambda)^2
\label{eq:linesig}
\end{equation}
where $P(\lambda)$ is the line profile and $P_0(\lambda)$ is the first moment of $P(\lambda)$. Using the line dispersion as the line width estimator generally provides better black hole mass estimates than using the full-width at half maximum \citep[e.g.,][]{Peterson2004,Dalla2020}. Given the low SNR of our single-epoch spectra, we measure the line dispersion from the mean spectra instead of the rms spectra commonly used in RM studies. Previous studies show that the mean spectra line dispersion reasonably approximates the rms spectra line dispersion \citep[e.g.,][]{Wang2020,Dalla2020}. We adopt a virial factor of $f=4.31 \pm 1.05$ from \citet{Grier2013Martini}. Table \ref{tab:result} gives the line dispersion and black hole mass estimates for our final sample. The black hole masses have an uncertainty of $\sim 0.4$ dex, which is mainly from the intrinsic scatter in the calibration of the virial factor \citep[e.g.,][]{Peterson2014}. We do not include detailed estimates of other measurement uncertainties since they are small relative to the uncertainty in the virial factor. 

Figure \ref{fig:RL} shows the Mg II $R-L$ relation for our final sample. Before this work, the largest Mg II lag sample was the 24 sources in the gold sample of \citet{Homayouni2020}. Our sample nearly doubles the total number of high-quality Mg II lags. We significantly extend the Mg II lag measurements toward higher redshifts and bridge the gap in luminosity between the \citet{Homayouni2020} sample and the high luminosity sources studied by \citet{Lira2018}, \citet{Czerny2019}, \citet{Zajacek2020} and \citet{Zajacek2021}, as illustrated in Figures \ref{fig:sample} and \ref{fig:RL}. Compared to the OzDES Y5 results from Y21, the full OzDES Mg II lag sample covers about a factor of five wider range in log($\lambda L_{3000}$), which allows us to better constrain the $R-L$ relation. 

We parameterize the $R-L$ relation as 
\begin{equation}
{\rm log}(R / \ltday) = \alpha \, {\rm log}[L / (10^{45}\,{\rm erg\cdot s^{-1}})] + \beta
\label{eq:rl}
\end{equation}
with an intrinsic scatter $\sigma_{\rm rl}$. We fit our final sample with the MCMC sampler \textsf{emcee} \citep{emcee} and get best-fit parameters $\alpha = 0.39 \pm 0.08$, $\beta = 2.07 \pm 0.04$ and $\sigma_{\rm rl} = 0.15^{+0.03}_{-0.02}$. Figure \ref{fig:RL} shows the best-fit $R-L$ relation and uncertainties. The slope is shallower than the $\sim 0.5$ slope of the H$\beta$ $R-L$ relations \citep[e.g.,][]{Bentz2013}, although they are marginally consistent at $1.5\sigma$. The slope is broadly consistent with the \citet{Homayouni2020} gold sample given their uncertainties, but our intrinsic scatter is significantly smaller than the $\sim 0.36$-dex intrinsic scatter in their sample. It is marginally larger than the intrinsic scatter $\sigma_{\rm rl} \sim 0.13$ dex of the H$\beta$ $R-L$ relation from \citet{Bentz2013} but within uncertainties, while it is smaller than $\sigma_{\rm rl} \sim 0.19$ dex of the H$\beta$ $R-L$ relation from \citet{Du2016} after they accounted for the effect of Eddington ratio. This indicates that the Mg II line may still have a tight $R-L$ relation despite its weaker response to the continuum variability \citep[e.g.,][]{Guo2020}.

For completeness, we also fit the $R-L$ relation including all Mg II lags from literature. The best-fit parameters are $\alpha = 0.3 \pm 0.05$, $\beta = 2.07 \pm 0.03$ and $\sigma_{\rm rl} = 0.24^{+0.03}_{-0.02}$. This gives a shallower slope and larger intrinsic scatter, although the intrinsic scatter is still smaller than that from \citet{Homayouni2020} and is close to the $\sim 0.26$-dex intrinsic scatter from \citet{Du2016} without considering the Eddington ratio. The lags from different works are selected using different criteria and have different definitions of uncertainties, which could bias the $R-L$ relation, so we adopt our fit to only the OzDES sample as the most robust Mg II $R-L$ relation.

\subsection{Effect of Iron Line Strength} \label{subsec:RFe}
\begin{figure}
\includegraphics[width=1.\linewidth]{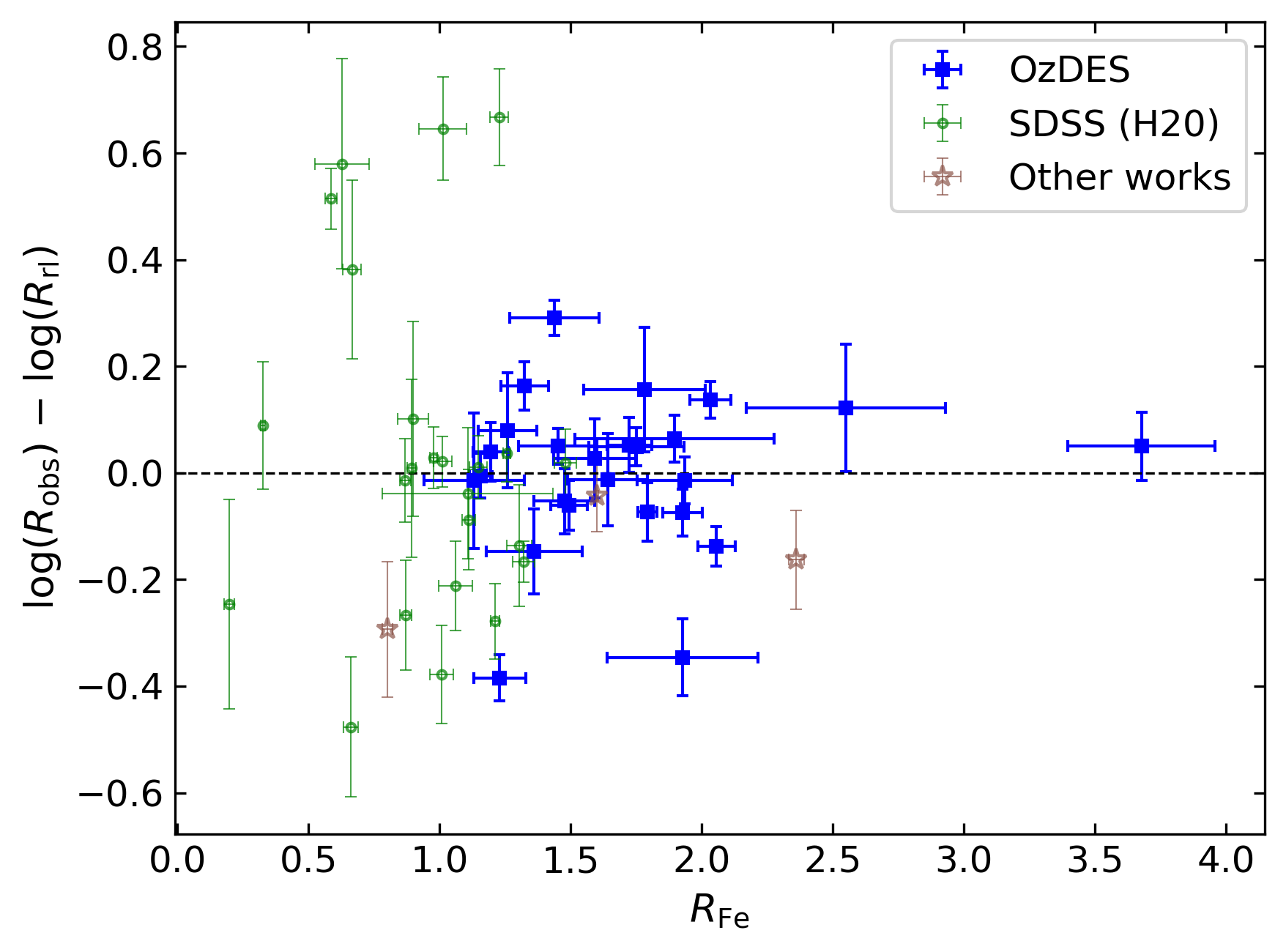}
\caption{Deviation of the observed lag $R_{\rm obs}$ from the prediction $R_{\rm rl}$ of the Mg II $R-L$ relation based on the OzDES sample versus the iron ratio $\mathcal{R}_{\rm Fe}$. The black dashed line is drawn where the observed lag falls exactly on the $R-L$ relation. The colored symbols have same meanings as Figure \ref{fig:RL}.}
\label{fig:difflag_RFe}
\end{figure}

Previous studies measured smaller H$\beta$ lags from AGN with higher Eddington ratios at fixed luminosities \citep[e.g.,][]{Du2016,Du2018,Dalla2020}. If also true of Mg II, it could explain the shallower slope of our Mg II $R-L$ relation. However, the Eddington ratios in previous studies depend on the virial black hole mass estimates and therefore depend on the lag measurements themselves. One independent indicator of the Eddington ratio is the strength of the iron emission \citep[e.g.,][]{Boroson2002,Negrete2018,Panda2019}. \citet{Du2019} reduced the intrinsic scatter of the $R-L$ relation by adding the ratio $\mathcal{R}_{\rm Fe, optical}$ of the optical iron line flux to the H$\beta$ line flux as a third parameter, while \citet{Khadka2022_Hbeta} found no significant reduction of the intrinsic scatter after including $\mathcal{R}_{\rm Fe, optical}$. This discrepancy could be because \citet{Khadka2022_Hbeta} constrained the $R-L$ relation and the cosmological parameters at the same time, while \citet{Du2019} fixed the cosmological parameters. \citet{Khadka2022} performed similar analysis with the ratio $\mathcal{R}_{\rm Fe}$ of the ultraviolet iron flux to the Mg II flux and found it had little impact on the Mg II $R-L$ relation. In this section we probe the effect of $\mathcal{R}_{\rm Fe}$ on our Mg II $R-L$ relation. 

For each spectroscopic epoch, we calculate the iron ratio as $\mathcal{R}_{\rm Fe} = F_{\rm Fe} / F_{\rm line}$, where $F_{\rm Fe}$ is the integrated iron flux over rest-frame 2250 - 2650 \AA\, and $F_{\rm line}$ is the Mg II line flux. We derive the mean $\langle \mathcal{R}_{\rm Fe} \rangle$ and the standard deviation $S_{\mathcal{R}_{\rm Fe}}$ over $N_{\rm e}$ epochs. The uncertainty of the mean is $\sigma_{\mathcal{R}_{\rm Fe}} = S_{\mathcal{R}_{\rm Fe}} / \sqrt{N_{\rm e}}$. Table \ref{tab:result} gives the iron ratio $\langle \mathcal{R}_{\rm Fe} \rangle$ and its uncertainty $\sigma_{\mathcal{R}_{\rm Fe}}$ for the AGN in our final sample. Figure \ref{fig:difflag_RFe} shows the deviation of the observed lags from our Mg II $R-L$ relation versus the iron ratio $\mathcal{R}_{\rm Fe}$. It also includes the AGN from previous Mg II RM studies, with the exception of \citet{Metzroth2006} and \citet{Lira2018} which did not report $\mathcal{R}_{\rm Fe}$ measurements. There is no significant correlation between the deviation from the $R-L$ relation and the iron ratio $\mathcal{R}_{\rm Fe}$. 

To quantify the potential improvement from the 2-parameter $R-L$ relation to a 3-parameter relation with $\mathcal{R}_{\rm Fe}$, we consider the two parameterizations studied in \citet{Khadka2022}:
\begin{equation}
{\rm log}(R / \ltday) = \alpha \, {\rm log}[L / (10^{45}\,{\rm erg\cdot s^{-1}})] + \beta + \gamma \mathcal{R}_{\rm Fe}
\label{eq:rl_RFe}
\end{equation}
and 
\begin{equation}
{\rm log}(R / \ltday) = \alpha \, {\rm log}[L / (10^{45}\,{\rm erg\cdot s^{-1}})] + \beta + \gamma\,{\rm log}(\mathcal{R}_{\rm Fe}).
\label{eq:rl_RFe}
\end{equation}
For the OzDES sample, we obtain an intrinsic scatter of $\sigma_{\rm rl} = 0.15^{+0.03}_{-0.02}$ for both parameterizations. When including the literature lags, both parameterizations give $\sigma_{\rm rl} = 0.22^{+0.03}_{-0.02}$. The intrinsic scatters of these 3-parameter correlations are close to the 2-parameter $R-L$ relation. These results indicate that the iron ratio $\mathcal{R}_{\rm Fe}$ has little impact on the Mg II $R-L$ relation, in agreement with the results of \citet{Khadka2022}.

\section{Summary} \label{sec:summary}
We use six years of photometry and spectroscopy from the OzDES RM project to measure Mg II lags. We calibrate the spectra using the pipeline developed by \citet{Hoormann2019} based on the DES photometry and estimate the calibration uncertainties using the Monte-Carlo approach from Y21 based on the F-stars observed in the OzDES fields. We use the algorithm from Y21 to model and subtract the continuum + iron emission around the Mg II line. We define quantitative criteria to select reliable measurements and verify the lag reliability using simulations based on both the DRW stochastic process and the observed lightcurves of AGN. Our major results are:

\begin{enumerate}
\item We obtain high quality Mg II lag measurements for 25 quasars. Seven quasars were presented in Y21, and their lags from the full 6-yr lightcurves are consistent with those from Y21 based on the first five years of data. Our sample substantially increases the number of Mg II lags and extends the $R-L$ relation to higher redshifts and luminosities.  
\\
\item Our sample provides a new constraint of the Mg II $R-L$ relation with a slope $\alpha = 0.39 \pm 0.08$, an intercept $\beta = 2.07 \pm 0.04$ and an intrinsic scatter $\sigma_{\rm rl} = 0.15^{+0.03}_{-0.02}$. The slope is consistent with the Mg II $R-L$ relation from \citet{Homayouni2020}, while it is shallower than the H$\beta$ $R-L$ relation based on local AGN. The intrinsic scatter is significantly smaller than that from \citet{Homayouni2020} and is close to that of the H$\beta$ $R-L$ relations, which makes it promising to apply our $R-L$ relation to large samples of single-epoch mass estimates and SMBH demographic studies at cosmic noon. 
\\
\item The residual from the $R-L$ relation has no significant correlation with the relative iron strength $\mathcal{R}_{\rm Fe}$ as an indicator of the Eddington ratio. Adding $\mathcal{R}_{\rm Fe}$ as a third parameter does not reduce the intrinsic scatter of the $R-L$ relation. 
\end{enumerate}

Future RM campaigns based on wide-field photometric and spectroscopic surveys, such as the Rubin Observatory Legacy Survey of Space and Time \citep[LSST, e.g.,][]{LSSTSciBook}, the Black Hole Mapper in SDSS-V \citep[e.g.,][]{SDSSV} and the Time-Domain Extragalactic Survey (TiDES) with 4MOST \citep[e.g.,][]{TiDES}, should provide larger sample of lag measurements and better constraints of the $R-L$ relation. In addition to deeper observations with future facilities, the optimization of survey strategy is critical for successful RM campaigns. For example, \citet{Malik2022} used simulated lightcurves to study the effect of observational windows on the lag recovery in future surveys. They found that maximizing the length of the observing seasons is especially important for reducing the aliasing effects that significantly affect the lag measurements, although observing polar fields could be challenging due to the higher air-mass. The spectroscopic calibration is also critical for future RM studies, since the calibration uncertainty could contribute a significant fraction of the error budget for deep observations. There are three AGN in our final sample where the calibration uncertainty dominates the error budget. Calibration uncertainties will become increasingly important for deeper future surveys.

\section*{Acknowledgements}
ZY and CSK are supported by Chandra grant GO9-20084X. PM is grateful for support from the Radcliffe Institute for Advanced Study at Harvard University. PM also acknowledges support from the United States Department of Energy, Office of High Energy Physics under Award Number DE-SC-0011726. CSK is supported by NSF grants AST-1908570 and AST-1814440. Parts of this research were supported by the Australian Government through the Australian Research Council Laureate Fellowship FL180100168.

Funding for the DES Projects has been provided by the U.S. Department of Energy, the U.S. National Science Foundation, the Ministry of Science and Education of Spain, 
the Science and Technology Facilities Council of the United Kingdom, the Higher Education Funding Council for England, the National Center for Supercomputing 
Applications at the University of Illinois at Urbana-Champaign, the Kavli Institute of Cosmological Physics at the University of Chicago, 
the Center for Cosmology and Astro-Particle Physics at the Ohio State University,
the Mitchell Institute for Fundamental Physics and Astronomy at Texas A\&M University, Financiadora de Estudos e Projetos, 
Funda{\c c}{\~a}o Carlos Chagas Filho de Amparo {\`a} Pesquisa do Estado do Rio de Janeiro, Conselho Nacional de Desenvolvimento Cient{\'i}fico e Tecnol{\'o}gico and 
the Minist{\'e}rio da Ci{\^e}ncia, Tecnologia e Inova{\c c}{\~a}o, the Deutsche Forschungsgemeinschaft and the Collaborating Institutions in the Dark Energy Survey. 

The Collaborating Institutions are Argonne National Laboratory, the University of California at Santa Cruz, the University of Cambridge, Centro de Investigaciones Energ{\'e}ticas, 
Medioambientales y Tecnol{\'o}gicas-Madrid, the University of Chicago, University College London, the DES-Brazil Consortium, the University of Edinburgh, 
the Eidgen{\"o}ssische Technische Hochschule (ETH) Z{\"u}rich, 
Fermi National Accelerator Laboratory, the University of Illinois at Urbana-Champaign, the Institut de Ci{\`e}ncies de l'Espai (IEEC/CSIC), 
the Institut de F{\'i}sica d'Altes Energies, Lawrence Berkeley National Laboratory, the Ludwig-Maximilians Universit{\"a}t M{\"u}nchen and the associated Excellence Cluster Universe, 
the University of Michigan, NSF's NOIRLab, the University of Nottingham, The Ohio State University, the University of Pennsylvania, the University of Portsmouth, 
SLAC National Accelerator Laboratory, Stanford University, the University of Sussex, Texas A\&M University, and the OzDES Membership Consortium.

Based in part on observations at Cerro Tololo Inter-American Observatory at NSF's NOIRLab (NOIRLab Prop. ID 2012B-0001; PI: J. Frieman), which is managed by the Association of Universities for Research in Astronomy (AURA) under a cooperative agreement with the National Science Foundation.

The DES data management system is supported by the National Science Foundation under Grant Numbers AST-1138766 and AST-1536171.
The DES participants from Spanish institutions are partially supported by MICINN under grants ESP2017-89838, PGC2018-094773, PGC2018-102021, SEV-2016-0588, SEV-2016-0597, and MDM-2015-0509, some of which include ERDF funds from the European Union. IFAE is partially funded by the CERCA program of the Generalitat de Catalunya.
Research leading to these results has received funding from the European Research
Council under the European Union's Seventh Framework Program (FP7/2007-2013) including ERC grant agreements 240672, 291329, and 306478.
We  acknowledge support from the Brazilian Instituto Nacional de Ci\^encia
e Tecnologia (INCT) do e-Universo (CNPq grant 465376/2014-2).

This manuscript has been authored by Fermi Research Alliance, LLC under Contract No. DE-AC02-07CH11359 with the U.S. Department of Energy, Office of Science, Office of High Energy Physics.

\section*{Data Availability}
We present supplementary materials in the online journal:
\begin{enumerate}
\item  Figures of the single-epoch spectra in the same format as Figure \ref{fig:ironfit} for the final sample sources. These figures also include the co-added and rms spectra for each source.
\item Machine readable lightcurves for the final sample sources and sources that only pass one set of criteria.
\item Figures of the lightcurves and lag measurements in the same format as Figure \ref{fig:lag_examp} for the final sample sources.
\item Figures of the simulated lightcurves based on the observed lightcurves of local AGN (Section \ref{subsec:simreflc}). They are in the same format as Figure \ref{fig:simreflc_examp} but for other reference AGN and input lag. 
\end{enumerate}
The underlying DES and OzDES data are available in \citet{DESDR2} and \citet{Lidman2020}.




\bibliographystyle{mnras}
\bibliography{ref} 


\vspace{0.4cm}
\noindent \textbf{Affiliations}\\
$^{1}$Department of Astronomy, The Ohio State University, Columbus, Ohio 43210, USA\\
$^{2}$Center of Cosmology and Astro-Particle Physics, The Ohio State University, Columbus, Ohio, 43210, USA\\
$^{3}$Radcliffe Institute for Advanced Study, Harvard University, Cambridge, MA 02138, USA\\
$^{4}$School of Mathematics and Physics, University of Queensland, Brisbane, QLD 4072, Australia\\
$^{5}$ARC Centre of Excellence for All-sky Astrophysics (CAASTRO), 44 Rosehill Street Redfern, NSW 2016, Australia\\
$^{6}$Sydney Institute for Astronomy, School of Physics, A28, The University of Sydney, NSW 2006, Australia\\
$^{7}$Australian Astronomical Observatory, North Ryde, NSW 2113, Australia\\
$^{8}$Research School of Astronomy and Astrophysics, Australian National University, Canberra, ACT 2611, Australia\\
$^{9}$Laborat\'orio Interinstitucional de e-Astronomia - LIneA, Rua Gal. Jos\'e Cristino 77, Rio de Janeiro, RJ - 20921-400, Brazil\\
$^{10}$Fermi National Accelerator Laboratory, P. O. Box 500, Batavia, IL 60510, USA\\
$^{11}$Sorbonne Universit\'es, UPMC Univ Paris 06, UMR 7095, Institut d'Astrophysique de Paris, F-75014, Paris, France\\
$^{12}$CNRS, UMR 7095, Institut d'Astrophysique de Paris, F-75014, Paris, France\\
$^{13}$University Observatory, Faculty of Physics, Ludwig-Maximilians-Universit\"at, Scheinerstr. 1, 81679 Munich, Germany\\
$^{14}$Department of Physics \& Astronomy, University College London, Gower Street, London, WC1E 6BT, UK\\
$^{15}$Instituto de Astrofisica de Canarias, E-38205 La Laguna, Tenerife, Spain\\
$^{16}$Universidad de La Laguna, Dpto. Astrofísica, E-38206 La Laguna, Tenerife, Spain\\
$^{17}$INAF-Osservatorio Astronomico di Trieste, via G. B. Tiepolo 11, I-34143 Trieste, Italy\\
$^{18}$Department of Astronomy, University of Illinois at Urbana-Champaign, 1002 W. Green Street, Urbana, IL 61801, USA\\
$^{19}$Center for Astrophysical Surveys, National Center for Supercomputing Applications, 1205 West Clark St., Urbana, IL 61801, USA\\
$^{20}$Institut de F\'{\i}sica d'Altes Energies (IFAE), The Barcelona Institute of Science and Technology, Campus UAB, 08193 Bellaterra (Barcelona) Spain\\
$^{21}$Institute for Fundamental Physics of the Universe, Via Beirut 2, 34014 Trieste, Italy\\
$^{22}$Astronomy Unit, Department of Physics, University of Trieste, via Tiepolo 11, I-34131 Trieste, Italy\\
$^{23}$Hamburger Sternwarte, Universit\"{a}t Hamburg, Gojenbergsweg 112, 21029 Hamburg, Germany\\
$^{24}$Centro de Investigaciones Energ\'eticas, Medioambientales y Tecnol\'ogicas (CIEMAT), Avda. Complutense 40, E-28040 Madrid, Spain\\
$^{25}$Jet Propulsion Laboratory, California Institute of Technology, 4800 Oak Grove Dr., Pasadena, CA 91109, USA\\
$^{26}$Institute of Theoretical Astrophysics, University of Oslo. P.O. Box 1029 Blindern, NO-0315 Oslo, Norway\\
$^{27}$Instituto de Fisica Teorica UAM/CSIC, Universidad Autonoma de Madrid, 28049 Madrid, Spain\\
$^{28}$Department of Physics and Astronomy, University of Pennsylvania, Philadelphia, PA 19104, USA\\
$^{29}$Department of Astronomy, University of Michigan, Ann Arbor, MI 48109, USA\\
$^{30}$Department of Physics, University of Michigan, Ann Arbor, MI 48109, USA\\
$^{31}$Observat\'orio Nacional, Rua Gal. Jos\'e Cristino 77, Rio de Janeiro, RJ - 20921-400, Brazil\\
$^{32}$Santa Cruz Institute for Particle Physics, Santa Cruz, CA 95064, USA\\
$^{33}$Department of Physics, The Ohio State University, Columbus, OH 43210, USA\\
$^{34}$Center for Astrophysics $\vert$ Harvard \& Smithsonian, 60 Garden Street, Cambridge, MA 02138, USA\\
$^{35}$Lowell Observatory, 1400 Mars Hill Rd, Flagstaff, AZ 86001, USA\\
$^{36}$Instituci\'o Catalana de Recerca i Estudis Avan\c{c}ats, E-08010 Barcelona, Spain\\
$^{37}$Department of Physics, University of Surrey, Guilford, Surrey, UK\\
$^{38}$Institute of Astronomy, University of Cambridge, Madingley Road, Cambridge CB3 0HA, UK\\
$^{39}$Department of Astrophysical Sciences, Princeton University, Peyton Hall, Princeton, NJ 08544, USA\\
$^{40}$Department of Physics and Astronomy, Pevensey Building, University of Sussex, Brighton, BN1 9QH, UK\\
$^{41}$School of Physics and Astronomy, University of Southampton,  Southampton, SO17 1BJ, UK\\
$^{42}$Computer Science and Mathematics Division, Oak Ridge National Laboratory, Oak Ridge, TN 37831\\
$^{43}$Institute of Cosmology and Gravitation, University of Portsmouth, Portsmouth, PO1 3FX, UK\\
$^{44}$Cerro Tololo Inter-American Observatory, NSF's National Optical-Infrared Astronomy Research Laboratory, Casilla 603, La Serena, Chile\\
$^{45}$Lawrence Berkeley National Laboratory, 1 Cyclotron Road, Berkeley, CA 94720, USA\\







\bsp	
\label{lastpage}
\end{document}